\documentclass[%
reprint,
nofootinbib,
 amsmath,amssymb,
 aip,
]{revtex4-1}

\usepackage{multirow}
\usepackage[normalem]{ulem}

\usepackage{longtable}

\usepackage{threeparttable}

\usepackage{graphicx}% Include figure files
\usepackage{dcolumn}% Align table columns on decimal point
\usepackage{bm}% bold math
\usepackage{hyperref}% add hypertext capabilities
\usepackage{bm}
\usepackage[english]{babel}
\usepackage[utf8]{inputenc}
\usepackage{amsmath}
\usepackage{amsfonts}

\usepackage{natbib}

\usepackage{graphicx}
\usepackage[colorinlistoftodos]{todonotes}
\usepackage{lmodern}

\usepackage{epstopdf}
\usepackage{hyperref}

\usepackage{braket}

\usepackage{tikz}
\usepackage{pgfplots}
\usetikzlibrary{plotmarks}

\definecolor{darkgreen}{rgb}{0,0.7,0}
\definecolor{lightblue}{rgb}{0.7,0.7,1.0}

\pgfplotscreateplotcyclelist{mycolor}{%
blue,every mark/.append style={fill=blue!80!black},mark=*\\%
red,every mark/.append style={fill=red!80!black},mark=square*\\%
darkgreen,every mark/.append style={fill=darkgreen!80!black},mark=diamond*\\%
black,mark=star\\%
blue,every mark/.append style={fill=blue!80!black},mark=diamond*\\%
red,densely dashed,every mark/.append style={solid,fill=red!80!black},mark=*\\%
darkgreen,densely dashed,every mark/.append style={
solid,fill=darkgreen!80!black},mark=square*\\%
black,densely dashed,every mark/.append style={solid,fill=gray},mark=otimes*\\%
blue,densely dashed,mark=star,every mark/.append style=solid\\%
red,densely dashed,every mark/.append style={solid,fill=red!80!black},mark=diamond*\\%
}

\pgfplotsset{cycle list name={mycolor}}

\pgfplotsset{
/pgfplots/bar cycle list/.style={/pgfplots/cycle list={%
{blue,fill=blue,mark=none},%
{red,fill=red,mark=none},%
{darkgreen,fill=darkgreen,mark=none},%
{black,fill=black,mark=none},%
}
},
}

\def\beq{\begin{equation}}
\def\eeq{\end{equation}}

\def\bs{\begin{split}}
\def\es{\end{split}}

\def\bea{\begin{eqnarray}}
\def\eea{\end{eqnarray}}

\renewcommand{\vec}[1]{\bm{#1}}
\newcommand{\kernel}{k}

\date{\today}% It is always \today, today,
\begin{document}

\preprint{APS/123-QED}

%\title{CQML: Approaching CCSD(T) accuracy by quantum machine learning on multi-space information}
%\title{Hierarchical improvements of quantum machine learning with multi-level combination technique}
\title{
Boosting quantum machine learning models with multi-level combination technique:
Pople diagrams revisited}

\author{Peter Zaspel}
\affiliation{Department of Mathematics and Computer Science, University of Basel, Spiegelgasse 1, 4051 Basel, Switzerland}

\author{Bing Huang}
\affiliation{%
Institute of Physical Chemistry and National Center for Computational Design and Discovery of Novel Materials (MARVEL),
Department of Chemistry, University of Basel, Klingelbergstrasse 80, 4056 Basel, Switzerland}%

\author{Helmut Harbrecht}
\email{helmut.harbrecht@unibas.ch}
\affiliation{Department of Mathematics and Computer Science, University of Basel, Spiegelgasse 1, 4051 Basel, Switzerland}
%\affiliation{Departement Mathematik \& Informatik, University of Basel, Spiegelgasse 1, 4051 Basel, Switzerland}

\author{O. Anatole von Lilienfeld}
\email{anatole.vonlilienfeld@unibas.ch}
\affiliation{%
Institute of Physical Chemistry and National Center for Computational Design and Discovery of Novel Materials (MARVEL),
Department of Chemistry, University of Basel, Klingelbergstrasse 80, 4056 Basel, Switzerland}%

\date{\today}% It is always \today, today,

\begin{abstract}
Inspired by Pople diagrams popular in quantum chemistry, we introduce a hierarchical scheme, based on the multi-level combination (C) technique,
to combine various levels of approximations made when calculating molecular energies within quantum chemistry. 
When combined with quantum machine learning (QML) models, the  
resulting CQML model is a generalized unified recursive kernel ridge regression 
which exploits correlations implicitly encoded in training data comprised of multiple levels in multiple dimensions. 
Here, we have investigated up to three dimensions: Chemical space, basis set, and electron correlation treatment. 
Numerical results have been obtained for atomization energies of a set of $\sim$7'000 organic molecules with up to 7 atoms (not counting hydrogens) containing CHONFClS,
as well as for $\sim$6'000 constitutional isomers of C$_7$H$_{10}$O$_2$. 
CQML learning curves for atomization energies suggest a dramatic reduction in necessary training samples calculated with the most accurate and costly method.
In order to generate milli-second estimates of CCSD(T)/cc-pvdz atomization energies with prediction errors reaching chemical accuracy ($\sim$1 kcal/mol), 
the CQML model requires only $\sim$100 training instances at CCSD(T)/cc-pvdz level, rather than thousands within conventional QML, while more training molecules are required at lower levels.
Our results suggest a possibly favorable trade-off between various hierarchical approximations whose computational cost scales differently with electron number.
\end{abstract}

\maketitle

\section{Introduction}\label{sec:introduction}
Chemical compound space, the property space spanned by all possible chemical compounds, is unfathomably large due to its combinatorial nature~\cite{ChemicalSpace,mullard2017drug}. 
Exploring chemical space from first principles is desirable in the context of computational materials design~\cite{ceder1998predicting,ComputationalMaterialsDesign_MRS2006,fpdesign2014anatole} as well as to fundamentally deepen our understanding of chemistry~\cite{anatole-ijqc2013}.
Over the last couple of years overwhelming evidence has been collected indicating that quantum machine learning (QML) models, trained throughout chemical space, hold great promise to dramatically reduce the cost for predicting quantum properties, such as atomization energies of molecules, for arbitrary out-of-sample molecules~\cite{CM,ML4Polymers_Rampi2013,Montavon2013,DTNN2017,Alan_OLED2015,Felix2016,Sandip2016,MLFF2017,googlePaper2017,amon,CeriottiScienceUnified2017,NeuralMessagePassing,schutt2018schnet,FCHL}. 
The core idea of QML is to learn the implicit mapping from geometrical and compositional information encoded in nuclear charges and positions to corresponding electronic properties from a set of 
training molecules with precomputed properties at a specific level of theory.  
The knowledge thus obtained from training is then applied to molecules out-of-sample, i.e., molecules not in the training set.
Nowadays, QML is a well-established technique and has several supervised learning variants, including mainly neural network \cite{Behlers_descriptor,DTNN2017,Alan_OLED2015} and kernel ridge regression \cite{CM,soap_2013,GAP}.
Currently, most of the efforts towards QML in literature are devoted to developing more efficient molecular representations \cite{baml,amon,FCHL} and adapting machine learning models to a growing number of applications \cite{tm_nn,MLspectra,Alan_OLED2015}. 
Recent overviews on the field were published in Refs.~\cite{RaghusReview2016,QMLessayAnatole,HMMChap2018} and an entire issue in J. Chem. Phys. was recently devoted to the theme of "Data-enabled theoretical chemistry"\cite{dataTheoChem2018}. 
%%%%%%%%%%%%%%%%%%%%%%%%%%%%%%
%TO RECONCILE WITH REST OF INTRO

%%%%%%%%%%%%%%%%%%%%%%%%%%%%%%

This progress was made possible due to the advent of modern computers which enabled routine calculations of  electronic properties such as ground state energies for large training sets of medium-sized organic molecules~\cite{DataPaper2014,smith2017ani,ghahremanpour2018alexandria} using common density functional approximations~\cite{DFT,dft_challenge_2012}. 
While QML prediction errors have been converged to values smaller than DFT accuracy~\cite{googlePaper2017}, the predictive power of any QML model inherently hinges on the accuracy of the employed reference data used for training. 
However, while the latest machine learning models are now able to make rather accurate and yet efficient predictions, the time required to compute training samples for large datasets with chemical accuracy is still prohibitive.
More specifically, in order to routinely match the experimental uncertainty of thermochemistry, the highly coveted ``chemical accuracy'' of $\sim$1 kcal/mol, typical approximations made within density functional theory do not suffice, and computationally expensive theories, e.g., CCSD(T) in a large basis, have to be used even when dealing just with closed-shell molecules in relaxed geometries. 
Unfortunately, due to its substantially larger computational complexity, the routine generation of CCSD(T) numbers in large basis sets for thousands of training molecules remains prohibitive.

The hierarchies encoded in model chemistries, well established in quantum chemistry, can be used to exploit systematic trends in cancellation of errors among different levels of theory, as proposed and demonstrated by Pople and co-workers~\cite{pople1999nobel,MolecularElectronicStructureTheory}. 
Composite methods are based on these ideas~\cite{ReviewCompositeMethods2016}, and include, among many others, Gaussian-n theories~\cite{PopleG2,PopleG3,PopleG3m}, the Weizmann-n methods~\cite{Weizmann_n_1,Weizmann_n_2}, and complete basis set (CBS) methods~\cite{CBS1,CBS2,CBS3}. 
They can reach chemical accuracy at the computational cost of combinations of more efficient models. 
When it comes to chemical space, the Pople diagram is a two-dimensional display of the relationship of the size of {\em any} molecule and level of theory~\cite{PopleDiagram}. 
Pople diagrams can easily be extended to accommodate additional or other dimensions such as relativistic effects~\cite{PopleDiagramRelativisticBasissetMethod} or accuracy~\cite{KarplusPopleDiagram}.
In this study, we apply the idea of a Pople diagram to combine varying levels of theory in the training set of QML models
(See Fig.~1 for the general idea). 
More specifically, we apply the sparse-grid combination (C) technique to estimate the optimal balance among (i) electron correlation (HF, MP2, CCSD(T)), (ii) basis set size (sto-3g, 6-31g, cc-pvdz), and (iii) number of organic molecules.
We find that the resulting CQML models require substantially less training instances at the computationally most demanding target level of theory.

To showcase our new developments, we will discuss a series of multi-level and multi-space machine learning models, as well as results for molecules from the QM7b dataset~\cite{qm7b}. Using several levels in the space of electron correlation approximations already leads to a very strong improvement in the learning results, with respect to the amount of necessary training data at target accuracy. Further improvement is found by adding different levels of basis sets.

This paper is structured as follows:
Section~\ref{sec:theory} briefly introduces the CQML model, as well as the data sets used for training and testing. In Section~\ref{sec:results}, results of the CQML model are presented and discussed for 2D  and 3d CQML models.  Finally, Section~\ref{sec:Conclusions} summarizes the main-findings, draws general conclusions and presents an outlook. 
Section~\ref{sec:theory} provides detailed methodological information to facilitate reproducibility of our findings.

%Model chemistry is “an approximate but well-defined mathematical procedure of simulation”1 of chemical phenomena. As pointed out by Pople,1[pople1999nobel] there is a wide range of possible empiricism; model chemistry can even be ab initio (i.e, without parameters except for fundamental constants of physics). Several multilevel model chemistry methods, such as the Gaussian-n theories and their variants developed by Pople and co-workers,2-5[PopleG2,PopleG3,PopleG3m,PopleG3x] the related Weizmann-n methods,6,7[Weizmann_n_1,Weizmann_n_2] the complete basis set (CBS) family of methods by Petersson and co-workers,8-10[CBS1,CBS2,CBS3] the single-coefficient 11-15[SingleCoeff_1,SingleCoeff_2,SingleCoeff_3,SingleCoeff_4,SingleCoeff_5] and multicoefficient 14-24[SingleCoeff_4,SingleCoeff_5,MCCM_1,MCCM_2,MCCM_3,MCCM_4,MCCM_5,MCCM_6,MCCM_7,MCCM_8,MCCM_9] correlation methods (MCCMs) of our group, and the recent multilevel methods of Hu and co-workers,25,26[Hu_MultLev_1,Hu_MultLev_2] have been developed and validated for covalent interactions as required for application to thermochemistry (heats of formation, atomization energies, etc.) and kinetics (barrier heights). 

%\section{Multi-Level Combination Technique Quantum Machine Learning (CQML) model}

\section{Computational Details}
\subsection{Datasets}
Two datasets were used for proof of principle: QM7b\cite{qm7b} and 6k constitutional isomers\cite{gdb9_data} (dubbed `CI9'), both are subsets of GDB-17 universe\cite{gdb17,gdb17_anie}. QM7b is composed of molecules with up to 7 heavy atoms, including C, N, O, S and Cl (H not counted), totaling 7211 molecules. Molecules in CI9 correspond to 6095 constitutional isomers of C$_7$H$_{10}$O$_2$.

For QM7b molecules, geometries were first optimized at the level of B3LYP/6-31g(d) using Gaussian 09~\cite{Gaussian09}, 
then single point calculations were calculated using three levels of theory (HF, MP2, CCSD(T)) and three basis sets (sto-3g, 6-31g and cc-pvdz) using Molpro ~\cite{MOLPRO}, resulting in 9 single point energies per molecule. 

For the CI9 molecules, three different methods were used: PM7, B3LYP/6-31g(2df,p) and G4MP2.
Relaxed geometries and energies were retrieved directly from reference~\cite{gdb9_data} for the latter two methods, 
while PM7 relaxed geometries and energies were obtained using MOPAC2016.~\cite{MOPAC}

\subsection{QML details}
We used both, the sorted Coulomb matrix~\cite{RuppPRL2012,AssessmentMLJCTC2013} and SLATM~\cite{amon} for modeling the CI9 data set, while SLATM~\cite{amon} only was used for QM7b.
Though slightly better performing representations have been published previously, 
such as SOAP~\cite{bartok2017sciadv,SOAP2018}, aSLATM~\cite{amon} or FCHL~\cite{FCHL}, 
comparison between CM and SLATM results indicates that trends are stable and that the conclusions drawn are independent of
choice of representation. 
As kernel-functions, we have always chosen the Laplace kernel $e^\frac{-\|\vec{R}_q-\vec{R}_i\|_1}{\sigma}$ with $\sigma$ being a hyper-parameter.
The hyper-parameter $\sigma$ was optimized manually and converged to $\sigma=400$. 
Furthermore we use a Lavrentiev regularization of size $10^{-10}$.
All presented errors are mean absolute error (MAE) comparing the prediction by the CQML method with the \textit{true} solution of the target theory level. 
The MAE is computed as out-of-sample error over 200 randomly chosen molecules that are not part of the training data set. 
These results are averaged over 20 training runs.
Note that we randomly choose the $N_{\ell_{\rm M} = 0}$ training molecules on the lowest level, 
while randomly selecting subsets of them on higher levels. 
This sequence of drawing ensures the nestedness of all the training samples.

\section{Theory} \label{sec:theory}
In this section, we start by reviewing systematic error cancellation, composite methods, the CQML approach, 
kernel ridge regression based QML and $\Delta$-ML \cite{DeltaPaper2015}, as well as two-, and $n$-dimensional CQML.

%\subsection{Composite methods}

\subsection{From Pople diagrams to CQML}
%In mathematics, a telescoping series is a series whose partial sums eventually only have a fixed number of terms after cancellation.~\cite{real_analysis} The cancellation technique, with part of each term cancelling with part of the next term, is known as the method of differences. 
%In spite of its simplicity, the concept of ``error cancellation'' is rather appealing in many fields of research, including computational chemistry. 

\begin{figure*}
\centering 
\includegraphics[scale=1.0]{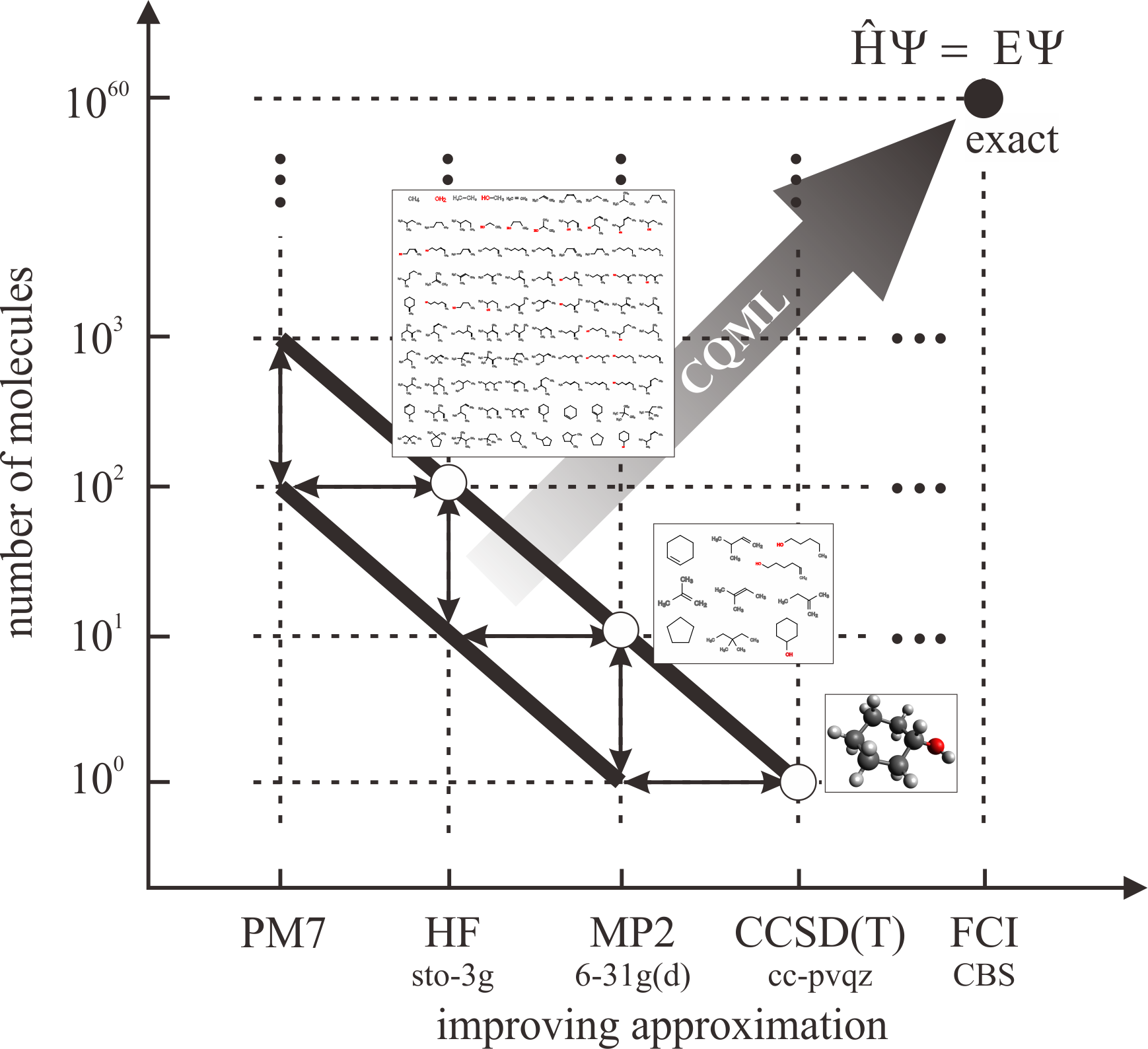}
\caption{\label{fig:pople_diagram} 
Adaptation of Pople diagram involving various levels of theory (abscissa) and molecular spaces (ordinate). 
The wide arrow indicates how to best approximate highly accurate solutions (solid black circle) of Schr\"odinger's
equation by combining ever improving levels of theory with an 
exponentially decreasing number of molecules used for training of machine learning models. 
Qualitative estimates of constant cost-benefit ratios (bold diagonals) correspond to Pareto-optimal solutions
which can be sampled using the CQML approach presented herewithin.
For example, training data consisting of 1 CCSD(T)/cc-pvqz, 4 MP2/6-31g(d), and 8 HF/sto-3g
calculation results can be cheaper {\em and} more valuable than 3 CCSD(T)/cc-pvqz results.
Two-sided arrows indicate bridges in chemical and method space.
}
\end{figure*}

Telescoping series, as a means to systematic convergence of error cancellation, are a well established mathematical tool.
In short, if $a_{n}$ is a sequence of numbers, 
then
\begin{equation}
\sum _{{\ell=1}}^{N}\left(a_{\ell}-a_{{\ell-1}}\right)= a_{N}-a_{{0}},
\end{equation}
and if we define $\Delta^{\ell}_{\ell -1} = a_{\ell }-a_{\ell-1}$ and $a_{0} = 0$, one has
\begin{equation} \label{eq:ec}
a_N = a_0 + \sum _{{\ell=1}}^{N} \Delta_{\ell-1}^{\ell}.
\end{equation}

Error cancellation is also at the root of many common practices in theoretical chemistry. 
Most notable are composite methods~\cite{PopleG2,PopleG3,PopleG3m,Weizmann_n_1,Weizmann_n_2,CBS1,CBS2,CBS3,griebel2008bossanova,griebel2014bond,chinnamsetty2018adaptive}, recently reviewed in Ref.~\cite{composite_review_karton2016}, which correspond to computational protocols which combine various quantum chemical approximations such that high accuracy (frequently chemical accuracy, i.e.~$\sim$ 1 kcal/mol) is achieved for thermodynamic quantities (e.g., atomization enthalpies). 
Typically,  they combine the results of a high level of theory with a small basis set with methods that employ lower levels of theory with larger basis sets. 
Importantly, they impose a computationally much reduced burden when compared to brute-force convergence in basis set size and electron excitations.
For example, an extensively used composite method called Gaussian-2 (G2),~\cite{G2} approximates the energy as 
(starting from a geometry optimized at MP2/6-31g(d) level)
\begin{equation}
E_{\rm true} \approx E^{\mathrm{G2}} := E_{\mathrm{QCISD(T)/6-311g(d)}} + \Delta_1 + \Delta_2 + \Delta_3,
\end{equation}
where further correction terms have been neglected. 
Note that here and throughout, we denote approximations and reference results by upper and lower indices, respectively. 
The individual terms read,
\begin{equation}
\begin{split}
\Delta_1 =& E_{\mathrm{MP4/6-311g(2df,p)}} - E_{\mathrm{MP4/6-311g(d)}},\\
\Delta_2 =& E_{\mathrm{MP4/6-311+g(d,p)}} - E_{\mathrm{MP4/6-311g(d)}},\\
\Delta_3 =& E_{\mathrm{MP2/6-311+g(3df,2p)}} + E_{\mathrm{MP2/6-311g(d)}} \\
& - E_{\mathrm{MP2/6-311g(2df,p)}} - E_{\mathrm{MP2/6-311g+(d,p)}}
\end{split}
\end{equation}
with $\Delta_1$ accounting for the effect of adding the polarization functions, $\Delta_2$ correcting for the diffuse functions and $\Delta_3$ correcting for the larger basis set as well as preventing contributions from being counted twice in  $\Delta_1$ and $\Delta_2$, respectively. 

Note that the formalism of the composite method corresponds to a sophisticated extension of the telescoping series in 
Equation (\ref{eq:ec}).  One could also simply rewrite (\ref{eq:ec}) as,
\begin{equation} \label{eq:ec_ML0}
\begin{split}
E_{\mathrm{CCSD(T)}} =& E_{\mathrm{HF}} + \Delta_{\mathrm{HF}}^{\mathrm{MP2}} + \Delta_{\mathrm{MP2}}^{\mathrm{CCSD(T)}},
\end{split}
\end{equation}
with all terms obtained for some large basis set. 
The problem then reduces to define efficient yet accurate estimates of the $\Delta$s. 
Here, we introduce the methodology to solve this problem through generalization 
of the $\Delta$-ML approach~\cite{DeltaPaper2015} in the form of CQML.

\subsection{The CQML approach}
\label{sec:mlctqml}

To exploit varying levels of theory in order to improve prediction accuracy, and thereby reduce the number of necessary costly training instances  
some of us previously introduced the $\Delta$-ML approach \cite{DeltaPaper2015}. It uses reference data calculated from a computationally efficient but inaccurate method as a baseline and estimates the difference to a more expensive but accurate target level of theory. Numerical results for organic molecules indicated that given an appropriately chosen baseline method, it is possible to achieve orders of magnitude reduction in training set size when compared to traditional QML approaches.
Many other studies have already shown the usefulness and applicability of the $\Delta$-ML 
approach~\cite{MLcrystals_Felix2015,Nick2016GA,raghu2015excitation,MLatoms_2015,Yang2018,dral2017,Simm2018,schmitz2018,bartok2017sciadv}. 
Alternatively, efforts have been made towards training set size reduction based on training set optimization \cite{Nick2016GA,amon,smith2018less,gubaev2018machine}
or improvements in the representations~\cite{baml,RuppsMBTR2017, CeriottiScienceUnified2017,FCHL}.
To the best of our knowledge, no conceptual improvements or generalizations of the $\Delta$-ML approach have been proposed so far. 

\begin{figure}[t]
\includegraphics[scale=0.2]{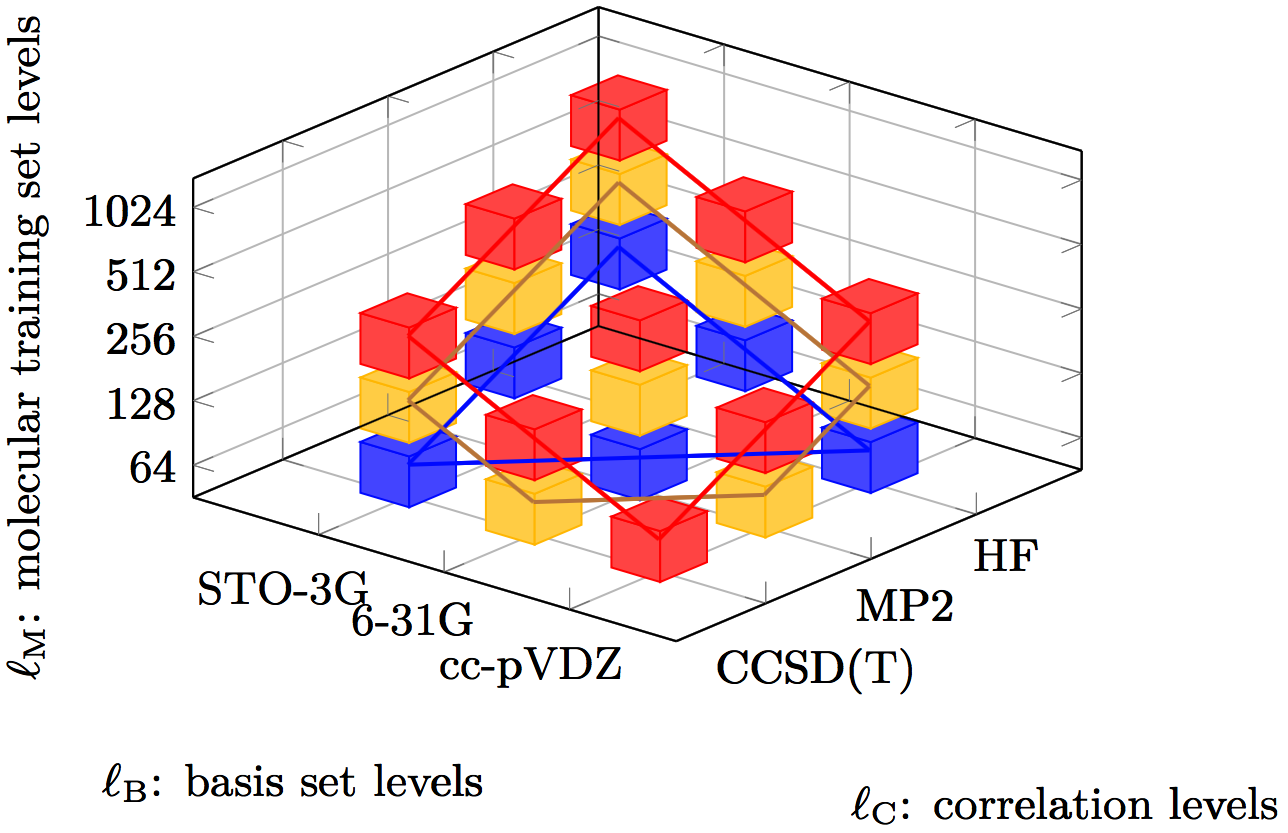}
\caption{\label{fig:ctSubspaces} 
The 3D CQML approach combines multiple levels in the spaces of electron correlation, basis sets, and training molecules. 
%\textit{Left:} Example of combination technique standard choices of subspaces. \textit{Right:} 
%Our adapted choice of subspaces for the 3D CQML model.}
}
\end{figure}

In this work, we generalize the core ideas of $\Delta$-ML \cite{DeltaPaper2015} to arrive at a multi-level combination technique QML (CQML) approach. 
CQML is a unified kernel ridge regression machine learning model incorporating training data from several spaces and levels of information. 
As proposed by e.g.~John Pople \cite{PopleDiagram,pople1999nobel}, we distinguish between 
\begin{enumerate}
\item the \textit{space of electron correlation} (e.g.~MP2) and
\item the \textit{space of basis set} (e.g.~6-31g), and we also add
\item the \textit{space of training molecules} (e.g.~some training set drawn from QM9~\cite{DataPaper2014}) 
 as third degree of freedom which can easily be exploited through machine learning models. 
\end{enumerate}
We call a specific choice of training information, e.g~Hartree-Fock calculations on a 6-31g basis set done for 256 molecules, a \textit{subspace}. 
Within each space, we assume a multi-level hierarchy of growing accuracy and computational complexity. 
E.g.~in electron correlation and basis set space, one commonly expects that
the degree of approximative nature decays systematically as one goes from HF to  MP2 to CCSD(T), from sto-3g to 6-31g to cc-pvdz, respectively. 
In chemical space, it is less obvious how to establish a hierarchy of accuracy. 
For the purpose of our approach, we rely on the well established tenet in statistical learning that 
the predictive accuracy for out-of-sample increases systematically with training set 
size~\cite{RasmussenWilliams}, which is applicable to chemical space and quantum chemistry as demonstrated first in 2012~\cite{RuppPRL2012}.
This finding has by now been confirmed and reproduced within multiple studies for various quantum properties and system 
classes~\cite{RaghusReview2016,QMLessayAnatole}. 
As such, and when drawing training molecules at random, we can consider their number made available to training (e.g.~$N=16, 32, 64\, \ldots$) 
as the chemical space equivalent to the space of theory (e.g.~HF/MP2/CCSD(T)) or basis set (e.g.~sto-3g/6-31g/cc-pvdz). 
Generally speaking, a CQML model built on low levels of theories / basis sets / small number of training molecules, 
will result in a model with low accuracy and easily accessible training data. 
Conversely, including more levels in each dimension, the resulting CQML model will become increasingly 
more accurate, requiring, however, also access to ever more valuable training data. 
Figure~1 exemplifies these ideas for various levels of electron correlation, basis sets, and molecular training set sizes.  

The \textit{sparse grid combination technique} known for high-dimensional approximation\cite{BG,GSZ,GH2,GH3,GH4,GK1,GK2,HGC,PF,REI,Ruettgers} and quadrature / uncertainty quantification\cite{HPS2,HPS4} %GHP,BGRZ,
in numerical analysis corresponds to a rigorous means 
to generate QML models constructed on a combination of sets of different subspaces. 
The general idea is to combine the subspaces such that only very few very expensive training samples are needed at target accuracy (e.g.~CCSD(T) for cc-pvdz at high sample count), some less expensive subspaces with higher training sample count are needed, and so on.
% in order to build a combined machine learning model at the target accuracy. 
Figure~\ref{fig:ctSubspaces} outlines a choice of subspaces by a modified sparse grid combination technique. 
Here, each subspace is represented by a colored cube.

In this work, we will first generalize the aforementioned $\Delta$-ML approach to a multi-level approach that incorporates the space of theories, basis sets, and training molecules. 
The CQML approach differs from existing multi-fidelity machine learning models~\cite{pilania2017multi} in that it is (a) generalized to multiple dimensions, 
and (b) does not unite the various spaces within one kernel matrix, but rather through a series of independently trained kernels.
%Based on this development, we will introduce the full sparse grid combination technique to quantum machine learning leading to our new CQML approach. 
While the CQML approach accounts for an arbitrary number of information spaces, for the sake of brevity and without any loss of generality, 
we restrict ourselves only to the three spaces discussed above.

\subsection{Kernel ridge regression and the $\Delta$-ML approach}\label{sec:methods_review}
In order to properly discuss CQML, we first need to briefly recall the principal idea of the established kernel ridge regression based QML models. 
With $\vec{R}$ (some) representation of a molecule, we denote by $E_\ell(\vec{R})$ the ML based approximation of the electronic ground state property 
of that molecule at a certain level of theory $l$. 
We train the ML model using $N$ training molecules $\vec{R}_i$ with $i=1,\ldots , N$ 
with corresponding reference energies at the corresponding specified level, $E_l^{\rm ref}(\vec{R})$. 
The objective is to predict energy $E_l^{\rm ref}$ for an out-of-sample \textit{query} molecule $\vec{R}_q$, neither part of training nor validation sets. 

The ML model $E_\ell$ within kernel ridge regression is then given by
$E_\ell^{\rm ref}(\vec{R}_q) \approx E_\ell(\vec{R}_q) := \sum_{i=1}^N \alpha_i^\ell \kernel(\vec{R}_q,\vec{R}_i)$, where $\kernel$ is an appropriate unit-less kernel function. 
For this study, we always choose the radial basis kernel function, $\exp[-\|\vec{R}_q-\vec{R}_i\|_1/\sigma]$ (Laplace) with length-scale $\sigma$.
Optimization of kernel function space could represent yet another potentially interesting dimension for future investigations.
As described in detail elsewhere~\cite{RasmussenWilliams,RaghusReview2016}, 
the coefficients $\alpha_i$ are obtained by solving the kernel matrix inversion problem
$\bm{\alpha} = ({\bm{K} + \lambda \bm{I}})^{-1} \bm{e}_\ell$ 
%$$P(\vec{R}_j) \approx \sum_{i=1}^N \alpha_i \kernel(\vec{R}_j,\vec{R}_i)\,,\quad j=1,\ldots,N$$ 
for given regularizer $\lambda$ and reference energy vector $\bm{e}_\ell$. 
Here, we use matrix-notation with capital and small case letters for matrices and vectors, respectively.
%$\vec{R}_i$ and $P(\vec{R}_i)$, cf.~\cite{} for more details on regression.

The $\Delta$-ML approach \cite{DeltaPaper2015} models the difference between a baseline and target level of  theory, e.g.~HF and MP2, respectively. 
Note, that we here have decided to adapt a slightly different notation in contrast to Ref.~\cite{DeltaPaper2015} in order to facilitate the generalization 
of the $\Delta$-ML to the CQML approach. 
Here, $P_{(b)}(\vec{R})$ and $P_{(t)}(\vec{R})$ represent the properties of interest computed at baseline and target level of theory, respectively. 
Note that within $\Delta$-ML, $P_{(b)}$ and $P_{(t)}$ it is not mandatory to estimate the same property, e.g.~it could be the ground state energy in the baseline theory and the enthalpy in the target theory. 
%The core idea of the $\Delta$-ML approach is to train the difference between the quantities $P^{(b)}$ and $P^{(t)}$ for the training samples on the respective levels. 
Hence, the $\Delta$-ML model prediction is given by
\begin{equation}
P_{(t)}(\vec{R}_q) := P_{(b)}(\vec{R}_q) + {\Delta_b^t}(\vec{R}_q)
\end{equation}
where
${\Delta_b^t}(\vec{R}_q) =  \sum_{i=1}^N \alpha_i \kernel\left(\vec{R}_q,\vec{R}_i\right)$.
%Again, the coefficients $\alpha_i$ are computed by solving a regression problem, namely
%$$P^{(t)}(\vec{R}_j) \approx P^{(b)}(\vec{R}_j) + \sum_{i=1}^N \alpha_i \kernel\left(\vec{R}_j,\vec{R}_i\right)$$
%with $j=1,\ldots, N$, for a given property $P$ evaluated on base and target level.
We emphasize that within the $\Delta$-ML model a potentially costly baseline evaluation of the query compound is still necessary when making a prediction.  
This differs from the CQML approach which recovers the original speed of QML by modeling even the baseline through a machine.
%However, in the following and without any loss of generality we will exclusively study only one property, namely the potential energy ground state energy, for the sake of simplicity. 

\subsection{Two-dimensional multi-level learning}\label{sec:mlqml}
The CQML approach generalizes the $\Delta$-ML model to several spaces {\em and} levels.
This is illustrated in Figure~\ref{fig:ctSubspaces} for three dimensions and levels 
%(spaces of basis set, theory, and molecules) and levels. 
%and three respective levels (sto-3g/6-31g/cc-pvdz, HF/MP2/CCSD(T), 64/128/256/512/1024) 
which we have also considered in this study ({\em vide infra}). 
To facilitate the discussion, we first discuss the adaptation of the Pople diagram in order 
to exemplify the general idea of the CQML approach for the simple case of only two dimensions. 
More specifically, we now consider the space of theory and training molecules. 
Thereafter, we will also discuss the generalization to three, as well as n-dimensional 
cases in Section~\ref{sec:mlctqml}.

Assuming $L$ levels of theory with running index $\ell=0, 1,\ldots,L-1$, for which the calculated energy increases in accuracy  $a$
(with respect to an experimentally yet unknown truth) and computational cost with growing theoretical complexity,
$a^{\ell+1} > a^{\ell} ,\; \forall \, \ell < L-1$.
%\begin{equation}\label{eq:errorDecay}a^{L} > a^{L-1} > \ldots > a^{l} > \ldots a^1 \,,\end{equation}
%where for each level $\ell$ the property $P^{(\ell)}$ can be computed with corresponding level of theory, 
%and where $e_{\ell_1}^{\ell_2}$ is the mean absolute (out of sample) error between $P^{(\ell_1)}$ and $P^{(\ell_2)}$ for a set of molecule samples. 
%The \textit{highest} level, i.e.~level $\ell=L$ corresponds to the target level of theory. 
%For example for $L=3$, one could combine HF, MP2 and CCSD(T) with CCSD(T) being the target theory.
%This is qualitatively correct for our example of HF, MP2 and CCSD(T).
% in the notion of $\Delta$-ML and HF and MP2 calculations are used to improve the results. 
Multi-level learning in two dimensions is performed as follows
\begin{itemize}
\item[(1)] on level $\ell=0$ compute reference energies $E^{\rm ref}_{\ell=0}$ for $N_{\ell=0}$ molecules
and train standard QML kernel ridge regression model to predict $E_{\ell = 0}$.
\item[(2)]  on level $\ell=1$ compute reference energies $E^{\rm ref}_{\ell = 1}$ for $N_{\ell = 1} < N_{\ell = 0}$ training molecules 
\item[(3)]  Still on level $\ell=1$, train a model of the difference between $E_0$ and $E_1^{\rm ref}$ for the $N_1$ molecules. 
\item[(4)] repeat recursively until target level $\ell=L-1$ is reached 
\end{itemize}
Note that while $N_{\ell}$ and $N_{\ell+1}$ molecules do not have to be identical, 
in this study all $N_{\ell+1}$ molecules are also part of the $N_{\ell}$ molecules out of convenience.

Formally, one can recursively define the intermediate multi-level 2D model $E_{\ell}$ 
for $\ell=0, 1, \ldots , L-1$, and built on the lowest level baseline ($\ell = 0$), as 
\begin{equation}\label{eq:2D}
E_{\ell}(\vec{R}_q) := E_{\ell-1}(\vec{R}_q) + \sum_{i}^{N_{\ell}} \alpha_i^{(\ell-1,\ell)} \kernel\left(\vec{R}_q,\vec{R}_i\right)\,,
\end{equation}
where we set $E^{0} \equiv 0$. 
For example, the CQML model which combines PM7 ($\ell = 0$), DFT ($\ell = 1$), and G4MP2 ($\ell = 2$) reads
\bea
%E^{\rm G4MP2}(\vec{R}_q) & =&  E^{\rm DFT}(\vec{R}_q) + \sum_i^{N_{\rm G4MP2}} \alpha_i^{({\rm DFT}, {\rm G4MP2})}k(\vec{R}_q,\vec{R}_i) \nonumber\\
E_{\rm 2}(\vec{R}_q) & =&  E_{\rm 1}(\vec{R}_q) + \sum_i^{N_{\rm 2}} \alpha_i^{({\rm 1}, {\rm 2})}k(\vec{R}_q,\vec{R}_i) \nonumber\\
{\rm where } && \nonumber \\
%E^{\rm DFT}(\vec{R}_q) & =& E^{\rm PM7}(\vec{R}_q) + \sum_j^{N_{\rm DFT}} \alpha_j^{({\rm PM7}, {\rm DFT})}k(\vec{R}_q,\vec{R}_j) \nonumber \\
E_{\rm 1}(\vec{R}_q) & =& E_{\rm 0}(\vec{R}_q) + \sum_j^{N_{\rm 1}} \alpha_j^{({\rm 0}, {\rm 1})}k(\vec{R}_q,\vec{R}_j) \nonumber \\
{\rm where} && \nonumber\\
%E^{\rm PM7}(\vec{R}_q) & =&  \sum_k^{N_{\rm PM7}} \alpha_k^{({\rm PM7})}k(\vec{R}_q,\vec{R}_k) \nonumber
E_{\rm 0}(\vec{R}_q) & =&  \sum_k^{N_{\rm 0}} \alpha_k^{({\rm 0})}k(\vec{R}_q,\vec{R}_k)\,, \nonumber
\eea
where the last term corresponds to the conventional direct QML model of the PM7 energy.
For numerical results obtained from this model, and their discussion {\em vide infra}. 
%However, the above notation also covers the case of skipping some levels at the beginning or at the end of the level hierarchy. 
%We will need this latter case in Section~\ref{sec:results}. 
To compute the coefficients $\alpha_i^{(\ell)}$, we solve the previously mentioned kernel ridge regression problem. 
%for $\ell=1,\ldots, N$  
%the regression problems to find $\alpha_i^{(\ell)}$ such that
%$$P^{(\ell)}(\vec{R}_j)\approx \M_{\ell_1}^{\ell-1}(\vec{R}_j) + \sum_{i=1}^{N_{\ell}} \alpha_i^{(\ell)} \kernel\left(\vec{R}_j,\vec{R}_i\right)\,,$$
%for all $j=1,\ldots, N_\ell$. 
%that are used in this preliminary version of our new method. Subspaces are shown for $L=3$ and a fixed basis set 6-31g.

%Namely, three terms involved in the r.h.s of equation (\ref{eq:ec_ML0}) can be accurately predicted by the means of kernel ridged regression approach, i.e., 
%\begin{eqnarray}
%%\begin{center}
%E_{\mathrm{HF}}(\vec{R}_q) = \sum_{i=1}^{N_1} \alpha_i^{\mathrm{HF}} k(\vec{R}_q, \vec{R}_i)\\
%\Delta_{\mathrm{HF}}^{\mathrm{MP2}}(\vec{R}_q) = \sum_{i=1}^{N_2} \alpha_i^{\mathrm{MP2}} k(\vec{R}_q, \vec{R}_i)\\
%\Delta_{\mathrm{MP2}}^{\mathrm{CCSD(T)}}(\vec{R}_q)  = \sum_{i=1}^{N_3} \alpha_i^{\mathrm{CCSD(T)}} k(\vec{R}_q, \vec{R}_i)
%%\end{center}
%\end{eqnarray}
%where $\vec{R}_q$ ($\vec{R}_i$) stands for the query (training) molecule, $\alpha$'s are the regression coefficients for training molecules, $k$ is the kernel matrix element measuring the similarity between two molecules, $N_1$, $N_2$ and $N_3$ run through all indexes of molecules used for training at the three corresponding levels of theory (note that the three sets of molecules can be chosen differently).
%

Let us briefly compare this approach to the conventional $\Delta$-ML models discussed before in Section~\ref{sec:methods_review}. 
In the single-level case, the resulting model $E_1$ is the direct conventional QML kernel ridge regression model. 
In the two-level case, the resulting model $E_2$ bears similarity with the $\Delta$-ML model, the major difference being that also the baseline is a machine. 
Thereby, it becomes possible to use different amounts of training information ($N_1$, $N_2$) on both levels. 
Nevertheless, if we chose the training molecules on the first and the second level identical and skipped regularization in the 
regression problem, $E_2$ and conventional direct QML would be identical.
And if we chose the training molecules on the first and the second level identical and built only one ML model (namely of the difference), 
$E_2^{\rm ref}$ and $\Delta$-ML would be identical.
$E_3$ and higher order approximations have, to the best of our knowledge, not yet been discussed in the literature.

Using above definition, we did not fix yet how to choose the amount of training samples on each level. 
This choice is based on the sparse grid combination technique \textcolor{green}{\cite{BG,GSZ,GH2,GH3,GH4,GK1,GK2,HGC,PF,REI,Ruettgers,HPS2,HPS4}}. %BGRZ,GHP,
Qualitatively, the combination technique implies to use many training samples on the lower levels of theory 
and to reduce the number of samples to very few samples on higher levels and the target level of theory. As we will see, the balance between the amount of training samples per level can be a point of optimization within our method.
%$\Rightarrow$ even better: the more levels, the higher the performance gain (cf.~Section~\ref{})
%First of all, it is necessary to define a \textit{measure of optimality} for such a balancing. In Section~\ref{sec:results}, we will discuss two \textcolor{red}{(?)} types of optimality measures. The first option is to measure the optimality of the method by the amount of expensive training samples that have to be evaluated on the highest level of theory. This rather simple optimality measure might already be a very good choice, since computing the training data on the target level of theory will usually dominate the overall computation time. The second option is to associate to each training sample (on the different levels) a \textit{cost} $c_i^{(\ell)}$. This might be the actual runtime of a quantum chemistry calculation or some generalized model. The optimality measure then corresponds to the sum over all costs on all levels. 
%\item second influence on optimal sample choice: the amount of error reduction on each level, cf.~\eqref{eq:errorDecay}
In Section~\ref{sec:results}, we discuss our choices of level balancing based on the sparse grid combination technique. 
These choices have been evaluated for different training data, and with respect to two possible optimality measures. 
Future work will deal with a more systematic assessment of how to tailor and optimize the
relative ratios of training molecules at each level and in each dimension.

\subsection{Three-dimensional multi-level learning}\label{sec:3Dcqml}
Extending Eq.~(\ref{eq:2D}) to more than two dimensions results in dimension-dependent levels. 
Table ~\ref{tab:Overview} provides an exemplifying overview for the three dimensions 
involving basis set (B), electron correlation (C), and molecular training set (M), with their corresponding
levels $\ell_B$, $\ell_C$ and $\ell_M$.

\begin{table}
\centering
\caption{
Exemplifying overview of levels in three dimensional multi-level learning for basis sets (B), electron correlation (C), and molecular training set (M). 
}
\begin{tabular}{|c|ccc|} \hline
  $level$ & 0 & 1 & 2 \\ \hline
  $\ell_{\rm C}$ &  HF & MP2 & CCSD(T) \\
  $\ell_{\rm B}$ &  sto-3g & 6-31g & cc-pvdz \\
  $\ell_{\rm M}$ &  $N_0$ & $N_1$ & $N_2$ \\\hline
\end{tabular}
\label{tab:Overview}
\end{table}

Thus, any given combination of levels can be specified as the ordered triplet $\ell$ of respective level indices, $\ell=(\ell_{\rm C}, \ell_{\rm B}, \ell_{\rm M})$.
For example, the combination CCSD(T)/cc-pvdz, $N_2$ is encoded by the triplet $\ell=(\ell_{\rm C} = 2, \ell_{\rm B} = 0, \ell_{\rm M} =1 ) = (2,0,1)$.
The corresponding CQML model is given by 
% \bea
% E_{201}(\vec{R}_q) 
% %&:= & \sum_{\ell_1+\ell_2+\ell_3 = 1} \sum_{i=1}^{N_{\ell_3}} \vec{\alpha}^{(\ell_1,\ell_2,N_{\ell_3})} k(\vec{R}_q,\vec{R}_i)\\
% %&-2 &\sum_{\ell_1+\ell_2+\ell_3 = 2} \sum_{i=1}^{N_{\ell_3}} \vec{\alpha}^{(\ell_1,\ell_2,N_{\ell_3})} k(\vec{R}_q,\vec{R}_i)\\
% %& + &\sum_{\ell_1+\ell_2+\ell_3 = 3} \sum_{i=1}^{N_{\ell_3}} \vec{\alpha}^{(\ell_1,\ell_2,N_{\ell_3})} k(\vec{R}_q,\vec{R}_i)\\
% & = & \sum_{\ell_{\rm C}+\ell_{\rm B}+\ell_{\rm M} = 1} E_{\ell_{\rm C}\ell_{\rm B}\ell_{\rm M}}(\vec{R}_q) \nonumber \\
% & - \, 2& \sum_{\ell_{\rm C}+\ell_{\rm B}+\ell_{\rm M} = 2} E_{\ell_{\rm C}\ell_{\rm B}\ell_{\rm M}}(\vec{R}_q) \nonumber\\
% & +&  \sum_{\ell_{\rm C}+\ell_{\rm B}+\ell_{\rm M} = 3} E_{\ell_{\rm C}\ell_{\rm B}\ell_{\rm M}}(\vec{R}_q)
% \label{eq:Example3D}
% \eea
$E_{(2,2,2)}$ and reads %obtain
\begin{equation} \label{eq:Example3D}
\begin{split}
E_{(2,2,2)}(\vec{R}_q) = & E_{(0,2,0)}(\vec{R}_q) - 2 E_{(0,1,0)}(\vec{R}_q)\\
& + E_{(1,1,0)}(\vec{R}_q) + E_{(0,1,1)}(\vec{R}_q)\\
& - 2 E_{(1,0,0)}(\vec{R}_q) + E_{(0,0,0)}(\vec{R}_q)\\
& - 2 E_{(0,0,1)}(\vec{R}_q) + E_{(2,0,0)}(\vec{R}_q) \\
& + E_{(1,0,1)}(\vec{R}_q)  + E_{(0,0,2)}(\vec{R}_q)
\end{split}
\end{equation}
with $\ell_{\rm C},\ell_{\rm B},\ell_{\rm M}=0,\ldots, 2$.
The reader is referred to Appendix~\ref{appendix} for the details of the derivation.

\subsection{n-dimensional multi-level learning}\label{sec:nmlqml}
Above, we discussed the low-dimensional multi-level method which profits from space of electron correlation, basis set, and molecular training set size. 
In order to extend this principle to even more spaces, we now generalize this approach following the lines of the sparse grid combination technique.
We introduce for $d$ spaces levels $\ell_1, \ldots , \ell_d$, which we collect together in the $d$-dimensional  \textit{multi-index} 
$\vec{\ell}=(\ell_1, \ldots , \ell_d)$. 
In the example of the previous section, $d=3$ and $\ell_1$ corresponds to $\ell_{\rm C}$,
$\ell_2$ corresponds to $\ell_{\rm B}$, and $\ell_3$ corresponds to $\ell_{\rm M}$. 
%Note that we here implicitly assume that the space of training data size will always be indexed by $d$. 
Following the notation that the last level index refers to molecular training set size, i.e.~$\ell_d = \ell_{\rm M}$, 
we define the energy $E^{\rm ref}_{(\vec{\ell})}$ given on a subspace $\vec{\ell}$, 
and the  QML model $E_{\vec{\ell}}$ for each subspace, 
\begin{equation}\label{eq:subspaceModel}
E_{\vec{\ell}}(\vec{R}_q) := \sum_{i=1}^{N_{\ell_d}} \alpha_i^{(\vec{\ell})} \kernel\left(\vec{R}_q,\vec{R}_i\right)\,.
\end{equation}
Computing the coefficients $\alpha_i^{(\vec{\ell})}$ for a fixed subspace $\vec{\ell}$ is done by solving the regression problem
\begin{equation}
E^{\rm ref}_{(\vec{\ell})}(\vec{R}_j)\approx \sum_{i=1}^{N_{\ell_d}} \alpha_i^{(\vec{\ell})} \kernel\left(\vec{R}_j,\vec{R}_i\right)
\end{equation}
for all $j=1,\ldots, N_{\ell_d}$. 

The generalized CQML machine learning model is then given as
\begin{equation}\label{eq:mlctqmlModel}
E_{\mathcal{I}}(\vec{R}_q) := \sum_{\vec{\ell}\in\mathcal{I}} \beta_{\vec{\ell}} \sum_{i=1}^{N_{\ell_d}} \alpha_i^{(\vec{\ell})} \kernel\left(\vec{R}_q,\vec{R}_i\right)\,.
\end{equation}

In fact, it is the combination of the machine learning models from \eqref{eq:subspaceModel} for different subspaces $\vec{\ell}$ that are collected in the \textit{index set} $\mathcal{I}$. The classical sparse grid combination technique proposes to use the index set
\bea 
\begin{split}
\mathcal{I} := \{ \vec{\ell}\in\mathbb{N}^d | \|\vec{\ell}\|_1 = & (L-1) - i, i\in\{0,\ldots,d-1\} \}
\end{split}
\eea
with $\|\vec{\ell}\|_1 := \sum_{s=1}^d \ell_s$. In the following, the coefficients $\beta_{\vec{\ell}}$ can always be evaluated as\cite{Ruettgers}
\bea 
\beta_{\vec{\ell}} := \sum_{\vec{z}\in\{0,1\}^d} (-1)^{\|\vec{z}\|_1} \chi_{_\mathcal{I}}(\vec{\ell}+\vec{z})\,.
\eea 
Here, the sum is to be understood in the sense that vector $\vec{z}$ of size $d$ takes all possible combinations of zeros and ones. Moreover we define the characteristic function $\chi_{_\mathcal{I}}$ of index set $\mathcal{I}$ by
\bea 
\chi_{_\mathcal{I}}(\vec{\ell}+\vec{z}) := \left\{\begin{array}{cl}1 & \mbox{if}\ (\vec{\ell}+\vec{z})\in\mathcal{I},\\ 0 & \mbox{else.}\end{array}\right.
\eea 
It is well-known \cite{} that the above choice of the index set $\mathcal{I}$ and coefficients $\beta_{\vec{\ell}}$ in $d=2$ is equivalent to the multi-level learning approach from Section~\ref{sec:mlqml}.%, that is, in $d=2$ the choice of subspaces corresponds to those given in Figure~\ref{fig:2dml}. 

For $d=3$ the above choice of index set $\mathcal{I}$ leads to the subspace choice in Eq.~\eqref{eq:Example3D}, exemplified with the spaces discussed in Section~\ref{sec:introduction}. Note, however, that this choice does not use any training data from the target subspace, here CCSD(T) calculations with a cc-pvdz basis set. In practice, it is preferable to include the corresponding subspaces with this accuracy to the training set, at least with a small training set size, in order to include the physics of the corresponding target accuracy. To this end, in $d=3$, we shift the index set $\mathcal{I}$ such that the subspace choice from Figure~\ref{fig:ctSubspaces} is achieved. %\textcolor{red}{Warning by PZ: This paragraph is incorrect. Figure~\ref{fig:ctSubspaces} only shows the shifted index set.} 
%For $d=3$ the above choice of index set $\mathcal{I}$ leads to the subspace choice shown in Figure~\ref{fig:ctSubspaces}, exemplified with the spaces discussed in Section~\ref{sec:introduction}. Note, however that this choice does not use any training data from the target subspace, here CCSD(T) calculations with a cc-pvdz basis set. In practice, it is preferable to include the corresponding subspaces with this accuracy to the training set, at least with a small training set size, in order to include the physics of the corresponding target accuracy. %To this end, in $d=3$, we shift the index set $\mathcal{I}$ such that the subspace choice from Figure~\ref{fig:ctSubspaces} is achieved. \textcolor{red}{Warning by PZ: This paragraph is incorrect. Figure~\ref{fig:ctSubspaces} only shows the shifted index set.} 

\section{Results and discussion}\label{sec:results}
Before entering the detailed discussion of our results, we now briefly discuss the use of learning curves as a measure
of machine learning model quality. 
Clearly, reporting a single out-of-sample error for any machine learning model is hardly meaningful: 
It is the very point of machine learning that models should improve with training set size.
Vapnik and co-workers discussed already in the nineties that prediction errors, i.e.~out-of-sample
estimates of statistically estimated functions, decay inversely with training set size $N$. 
More specifically, for kernel ridge regression models (used throughout this study),
the leading prediction error term was shown to be proportional to $a/N^b$, where $a$ and $b$
are proportionality constant and power law exponent, 
respectively~\cite{vapnik1994learningcurves,LearningCurvesMuller1996,vapnik2013book}.
In order to facilitate comparison among models, it is therefore recommended practice~\cite{QMLessayAnatole} 
to discuss the performance in terms of learning curves on log-log scales, i.e.~for 
prediction errors decaying linearly with training set size, i.e.~$\log({\rm Error}) = \log(a) - b \log(N)$. 
Saturation of errors  indicates failure to learn; and small off-sets and steep slopes indicate preferable
models. 

\begin{figure*}[t]
%\begin{tabular}{ccc}
%  \includegraphics[width=25mm]{DataPlot/a1u.pdf} &   \includegraphics[width=25mm]{DataPlot/a2u.pdf} &   \includegraphics[width=25mm]{DataPlot/a3u.pdf} \\
%  \includegraphics[width=25mm]{DataPlot/b1u.pdf} &   \includegraphics[width=25mm]{DataPlot/b2u.pdf} &   \includegraphics[width=25mm]{DataPlot/b3u.pdf} \end{tabular}
%\includegraphics[scale=0.4]{data.png}
%\includegraphics[scale=0.6]{s1_U.png}
%\includegraphics[scale=0.6]{s2_U.png}
\includegraphics[scale=0.6]{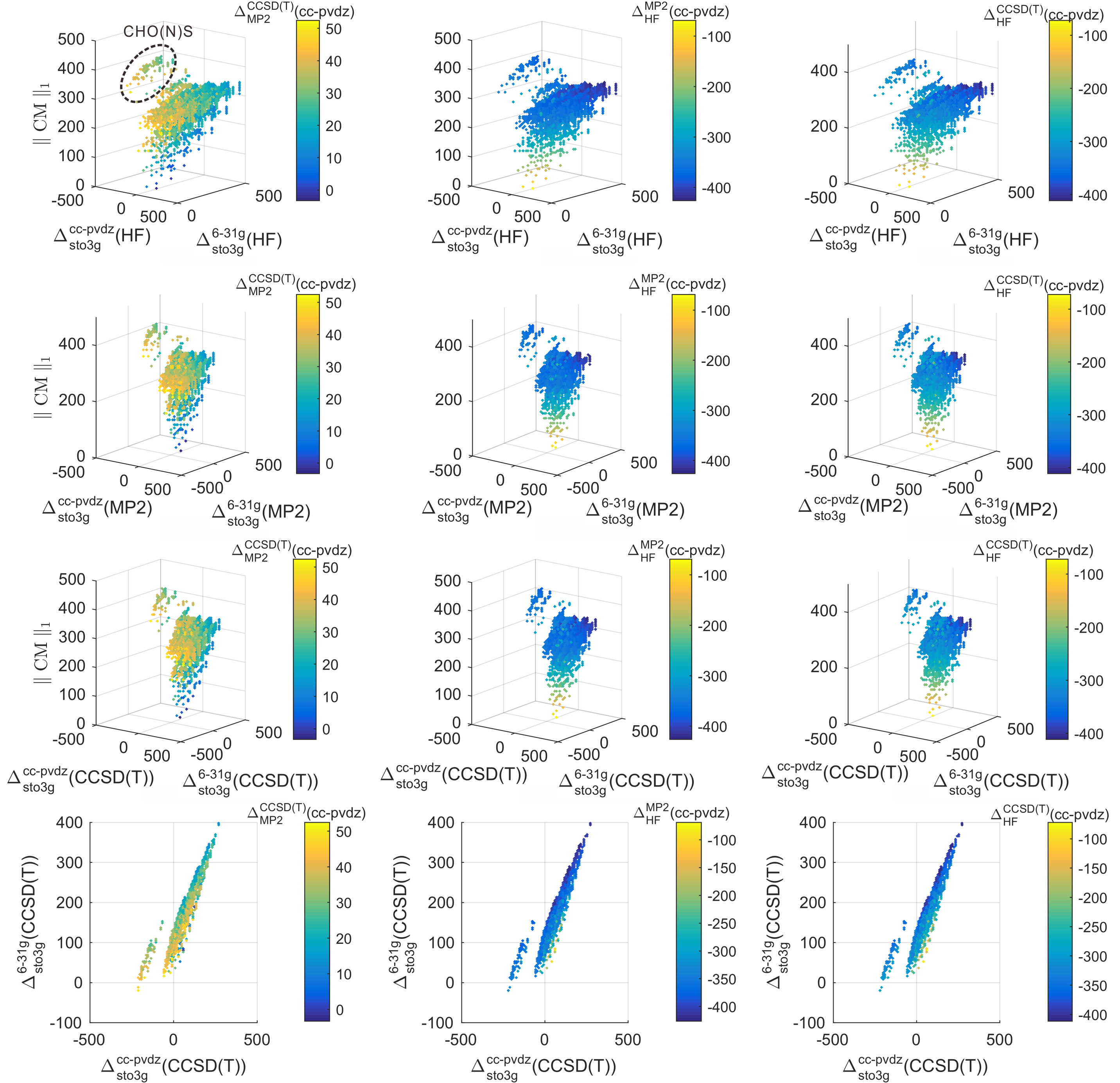}
\caption{
Scatter plots for QM7b. 
Size in chemical space as measured by $1$-norm of Coulomb matrix [a.u.] (i.e., $\| \mathrm{CM} \|_1$) vs.~energy differences [kcal/mol] due to 
various basis set size differences for HF (first row), MP2 (second row), and CCSD(T) (third row). 
The colour code corresponds to the atomization energy difference $\Delta$ [kcal/mol] between electron correlation models at cc-pvdz for MP2 vs.~CCSD(T) (left), HF vs.~MP2 (mid), and HF vs.~CCSD(T). 
In the upper leftmost panel, the brackets enclosing N indicate that nitrogen atoms may or may not be present. 
The bottom row corresponds to the 2D projection of the third row. 
}
\label{fig:corr}
\end{figure*}

\subsection{Data}
For all the $\sim$7'000 QM7b molecules~\cite{Montavon2013}, we have calculated total energies for all 
combinations among the various levels of correlation energies (HF, MP2, CCSD(T)) and basis set sizes (sto-3g, 6-31g, cc-pvdz).
Resulting effective atomization energies (see SI for the entire data set), are 
shown within scatter-plots in Fig.~\ref{fig:corr}.
%Depending on stoichiometry, degree of saturation, and size, the molecules spread out over the various levels and dimensions.
% added by s.b.
Depending on stoichiometry and size, the molecules spread out over the various levels and dimensions.
%Interestingly, various clusters form, corresponding to saturated and unsaturated sulphur containing molecules with nitrogen or oxygen.

\begin{figure*}[t]
\includegraphics[scale=0.4]{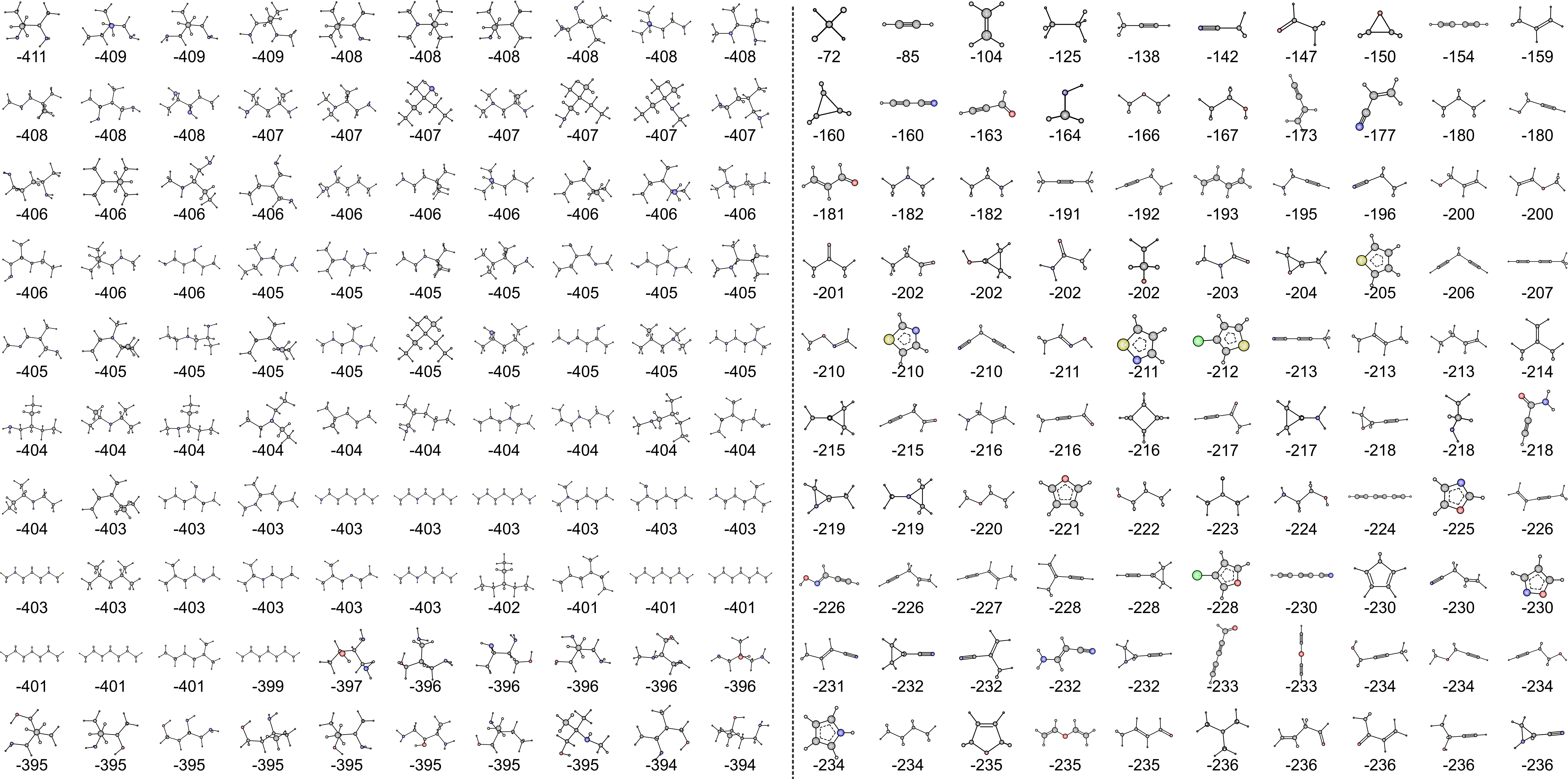}
\caption{
The two hundred QM7b molecules with largest (Left) and smallest (Right) electron correlation energy contributions to the atomization energy (CCSD(T) - HF within cc-pvdz basis [kcal/mol]), respectively.
See SI for the complete data set.  
White: H, gray: C, yellow: S, red: O, blue: N, green: Cl.
\label{fig:Molecules}
}
\end{figure*}

% added by s.b.
More specifically, molecules can be divided into two clusters: the one dominating the distribution is almost sulfur-free; while the other cluster of molecules, clearly separated from the majority, contains sulfur atoms (see bottom row in Fig.~\ref{fig:corr}). 
This pattern indicates that sto-3g and 6-31g are too small basis sets, and should not be used to describe S containing molecules. 
%for which typically at least split-valence basis set or more often diffuse and polarization basis functions also have to be used~\cite{}.
By comparing the three figures in each column of the first three rows in Figure~\ref{fig:corr}, one can see that the shape of distribution changes significantly upon introduction of electron correlation (going from HF treatment to the MP2). When going from MP2 to CCSD(T), however, the change in the distribution is barely noticeable. 

Considering the right hand panel in the third row in Fig.~\ref{fig:corr}, the color code corresponds exactly to the correlation energy contribution to the atomization energy, as estimated by CCSD(T) - HF within cc-pvdz basis. 
As one would expect, the larger the molecule, the more electron correlation energy is being contributed. 
The two hundred molecules with the largest and smallest correlation energy contribution to the atomization energy are on display in Fig.~\ref{fig:Molecules}. 
We note that molecules with high degree of saturation exhibit the largest amount of electron correlation in their atomization energy, while atomization energies of molecules with multiple double bonds, triple bonds, and aromatic moieties contain the least electron correlation energy.
This trend is to be expected because the electrons in unsaturated bonding patterns can contribute less to binding than in saturated species, thereby also decreasing their electron correlation energy contribution to binding. 

%Clearly, systematic trends can be seen by mere inspection. What is clear and to be expected is the trend of ranges: $| \Delta_{\mathrm{HF}}^{\mathrm{MP2}} | \gtrsim | \Delta_{\mathrm{HF}}^{\mathrm{CCSD(T)}} | > | \Delta_{\mathrm{MP2}}^{\mathrm{CCSD(T)}} |$ for any fixed basis set, as manifested by the fact that often MP2 can capture the essential part of but overestimate the correlation energy. 
%The spread within each cluster is dominated by differences in geometries. 
The reason for developing the CQML model is based on the hypothesis that it will systematically exploit all these underlying implicit correlations which are on display in these figures.

\begin{figure*}[t]
\includegraphics[scale=0.2]{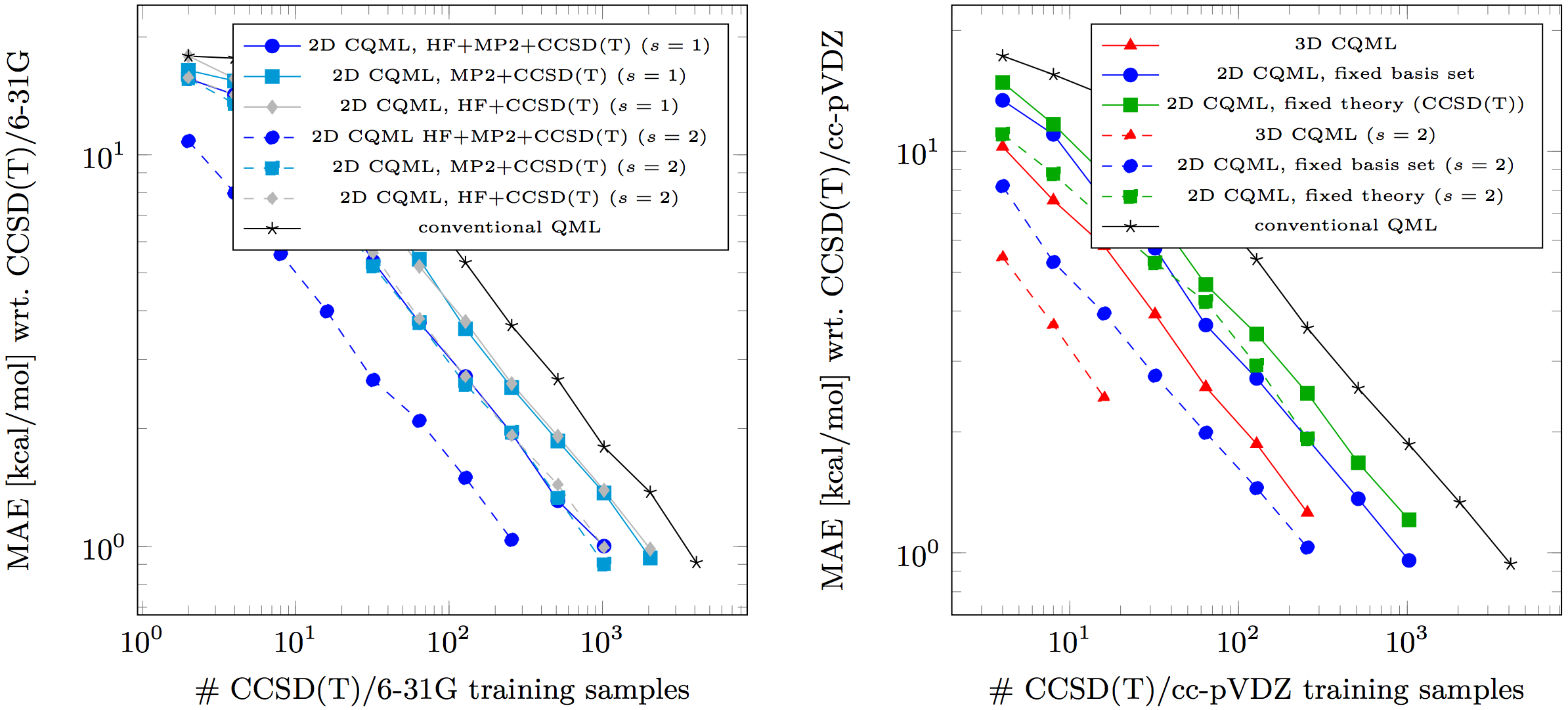}
\caption{
Prediction errors of CCSD(T) atomization energies in QM7b data set vs.~number of training molecules with CCSD(T) energies for various CQML models with level ratios
set by $s=1$ and $s=2$ (See table ~\ref{tab:ratios}). 
Left: 
2D-CQML at fixed basis set (6-31g) including 2 (MP2, CCSD(T))\textcolor{black}{/(HF,CCSD(T))}, and 3 levels of electron correlation treatment (HF, MP2, CCSD(T)).
Right: 
2D-CQML (green) at fixed electron correlation treatment (CCSD(T)) for 3 basis set sizes (sto-3g, 6-31g, cc-pvdz). 
3D-CQML (red) exploiting basis set size (sto-3g, 6-31g, cc-pvdz) {\em and} electron correlation treatment (HF, MP2, CCSD(T)). 
\label{fig:QM7b}
}
\end{figure*}

\subsection{2D results for QM7b}
\label{sec:results2d}
As a first test, we have investigated our QM7b derived data set for the two dimensional ($d = 2$) case of 
atomization energies at a fixed basis set (6-31g) for three levels of electron correlation, 
i.e.~HF ($\ell_{\rm C} = 0$), MP2 ($\ell_{\rm C} = 1$) and CCSD(T) ($\ell_{\rm C} = 2$). 
The second dimension corresponds to three variable molecular training data set sizes ($\ell_{\rm M} = 0, 1, 2$).
Their relative extent is fixed at ratios which are independent of absolute training set size.
In this study, we considered two such sets of ratios ($s=1$ and $s=2$) which reflect different
sample size increases for higher levels.
These ratios are  summarized in Table~\ref{tab:ratios}.
The number of training molecules $N_{\ell_{\rm M}}$ on each level of the CQML with $d=2$ 
as a function of training set size at the highest level $N_{\ell_{\rm M}=2}$ 
is thus given by $N_{\ell_{\rm M}} = r_{\ell_{\rm M}} \times N_{\ell_{\rm M}=2}$, where $r_{\ell_{\rm M}}$ is the ratio as displayed in Table~\ref{tab:ratios}.
Recall that all ML model results presented in this section have been obtained using
kernel ridge regression, a Laplacian kernel, and the SLATM~\cite{amon} representation.

\begin{table}
\centering
\caption{
Level-dependent ratios between training set sizes for the two sample size increases $s$ considered.  
$L$ is the total number of levels.
}
%\begin{tabular}{|d|c|ccccc|} \hline
\begin{tabular}{|c|ccc|} \hline
 $s$ & $r_{\ell_{\rm M} = L-1}$ & $r_{\ell_{\rm M} = L-2}$ & $r_{\ell_{\rm M} = L-3}$  \\ \hline
  1   &  1 & 2 & 4 \\
  2   &  1 & 4 & 16  \\ \hline
%$d$ &  $s$ & $r_{\ell_{\rm M} = 2}$ & $r_{\ell_{\rm M} = 1}$ & $r_{\ell_{\rm M} = 0}$ & & \\ \hline
% 2  &  1   &  1 & 2 & 4 & - & - \\
% 2  &  2   &  1 & 4 & 8 & - & - \\ \hline
% 3  &  1   &  1 & 2 & 4 & & \\
% 3  &  2   &  1 & 4 & 8 & & \\ \hline
\end{tabular}
\label{tab:ratios}
\end{table}

In Figure~\ref{fig:QM7b}, various learning curves for atomization energies, estimated according to Eq.~\eqref{eq:2D}, are shown. 
First of all, we note the rapid and systematic lowering for all CQML models as training set size increases.
The models exhibit differing off-sets, and similar slopes, in line with previous results for training-set optimization
experiments using ensembles of training sets within genetic optimization protocols~\cite{Nick2016GA}.  
The learning curves of conventional QML pass the chemical accuracy threshold ($\sim$1 kcal/mol) at 
$\sim 4'000$ training molecules calculated at target level, CCSD(T)/6-31g.
This learning curve has a slightly larger off-set with respect to the original SLATM benchmark results 
(see supplementary materials in Ref.~\cite{amons2017}) due to the use of 
(i) the Laplacian instead of a Gaussian kernel function, (ii) B3LYP rather than PBE0 geometries, 
and (iii) CCSD(T) rather than PBE0 energies.
%$N_{\ell_{\rm M}=2} 

Addition of MP2 reference energies of further molecules 
affords a systematic decrease in the learning off-set resulting 
in $\sim$2'000 and $\sim$1'000 CCSD(T) training molecules necessary to reach chemical accuracy
for $s = 1$ and $s = 2$, respectively.
The corresponding necessary MP2 training set sizes (not shown in the figure) amount to 
4'000 molecules for both $s$-values (see Table ~\ref{tab:ratios}). 
% added to s.b.2
\textcolor{black}{Slightly worse results are obtained by replacing MP2 reference
energies with HF energies.} 
% added by s.b.
\textcolor{black}{This result may seem puzzling, but is in full agreement with what we have found in~Figure~\ref{fig:corr}, i.e., the values of $\Delta_{\mathrm{MP2}}^{\mathrm{CCSD(T)}}$ and $\Delta_{\mathrm{HF}}^{\mathrm{MP2}}$ are of the same magnitude. This result also implies the possibility to optimize the levels of theory by minimizing the computational cost, meanwhile retaining the accuracy. }

Adding Hartree-Fock treatment for additional training molecules, we observe even further
improvement, reaching chemical accuracy already at $\sim$1'000 and $\sim$300 CCSD(T)
training molecules for $s = 1$ and $s = 2$, respectively.
According to the ratios in Table~\ref{tab:ratios},
the corresponding necessary MP2 and HF training set sizes (not shown in the figure) amount respectively to 
2'000 and 4'000 for $s = 1$, and to 1'200 and 2'400 for $s = 2$.

These results are very encouraging; they suggest that reductions by an order of magnitude are possible with respect 
to high-level reference numbers (from expensive computation or experiment) necessary to reach chemical accuracy. 
Effectively, the CQML model appears to exploit correlations inherent among the various approximation levels
that live within hierarchical spaces of theories.

\subsection{3D results for QM7b} \label{sec:results3d}
We have also studied the extension of the 2D-CQML model by a third dimension ($d=3$) which explicitly introduces the effect of basis set size on atomization energies.
More specifically, we have considered sto-3g ($\ell_{\rm B} = 0$) as our lowest level,
6-31g ($\ell_{\rm B} = 1$) as an intermediate size, and cc-pvdz ($\ell_{\rm B} = 2$) as the largest set. 
Obviously, larger basis set choices as well as additional levels with more subtle differences could have been included just as well. Here, we assume that the general trend and the conclusions drawn are not affect by the relatively modest size of the basis sets employed.

In Fig.~\ref{fig:QM7b}, we show corresponding learning curves of 2D-CQML models which connect the different  basis sets
according to Eq.~\eqref{eq:2D} with just one correlation energy model, CCSD(T).
In line with the behavior encountered above for the fixed basis set CQML models, a systematic improvement is 
found. The error approaches chemical accuracy already with $\sim$1'000 training examples with the largest basis used (cc-pvdz). 
Again, increasing the ratios between levels by going from $s = 1$ to $s = 2$ (see Table~\ref{tab:ratios}) 
leads to systematic lowering of the learning curve.

Finally, when combining multiple basis set and electron correlation levels into a single 3D-CQML model, 
obtained according to Eq.~\eqref{eq:Example3D}, %{\color{red} THIS EQUATION STILL NEEDS ATTENTION!},
the most favorable learning curves are obtained (See Fig.~\ref{fig:QM7b}).  
For $s=1$ and $s=2$, extrapolation indicates that chemical accuracy can be reached
with just 500 and 100 training instances at CCSD(T)/cc-pvdz level, respectively.
Note that the learning curves end already for relatively small training set sizes 
because the necessary number of molecules required at lower levels of theory 
rapidly reaches the maximal number of available molecules in QM7b. 
For example, for the $s=2$ case, 100 training molecules at the highest level combination would have 
required 100$\times$4$^4$ = 25,600 training molecules at the lowest level combination.
However, QM7b is comprised of only 7'211 molecules.
As such, this is an artefact of the finite size of QM7b, and we expect these learning curves
to further decay linearly when using larger data sets in the future.

Overall, these results amount to numerical evidence that it is beneficial to include
not only multiple levels but also multiple dimensions. 
The obvious consequence is that an additional substantial reduction in need for 
high-level reference numbers (from expensive computation or experiment) is possible through the 
use of CQML based exploitation of training data obtained for smaller basis sets and more approximate electron correlation models. 
We believe that this is possible because of inherent error-cancellation between 
various levels {\em and} dimensions.

\begin{figure}[t]
\includegraphics[scale=0.2]{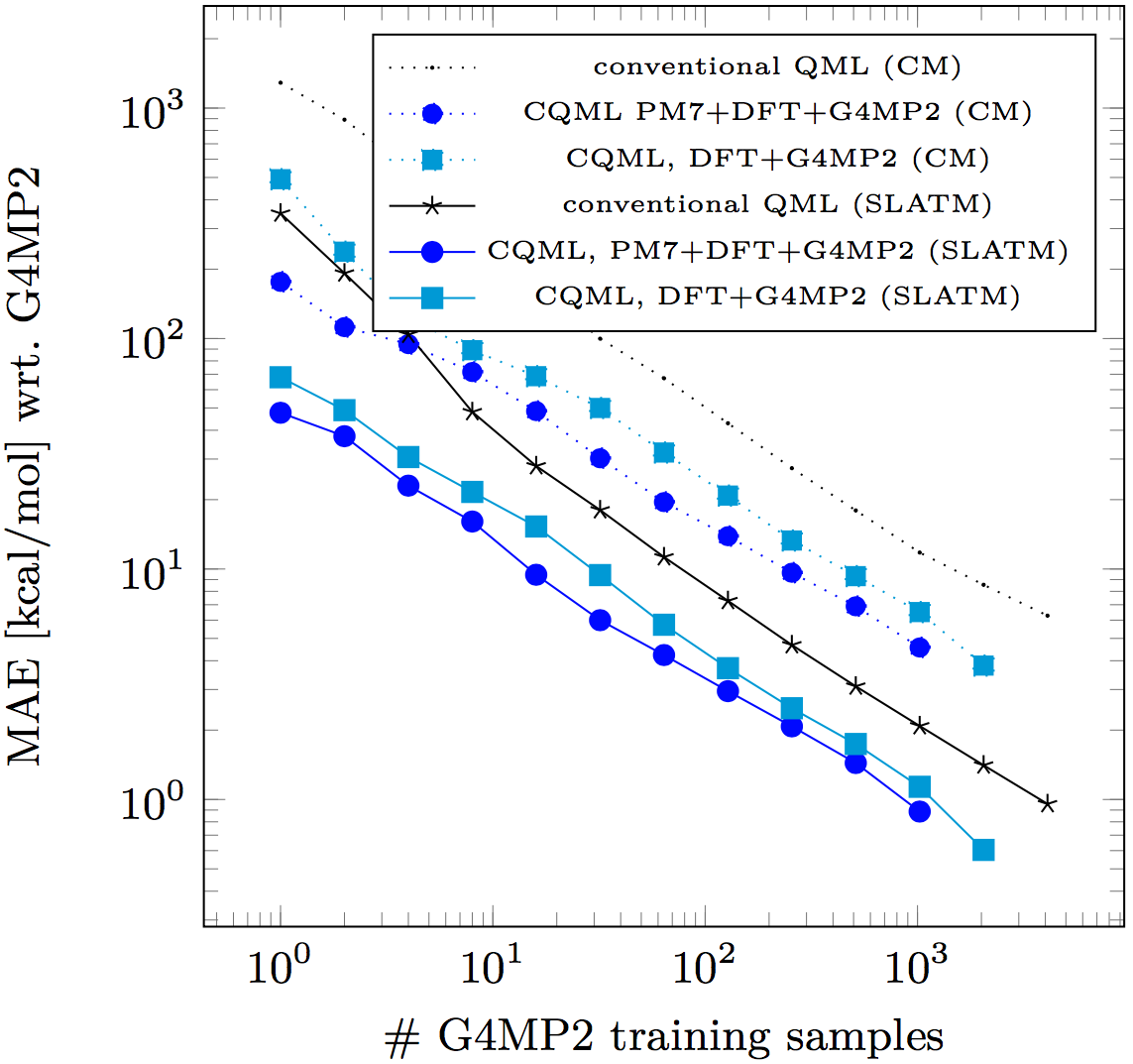}
\caption{Prediction errors of atomization energies in CI9 data set (consitutional isomers of C$_7$H$_{10}$O$_2$) vs.~number of training molecules with G4MP2 energies for various 2D-CQML models. 
Results differ by representation (SLATM vs.~CM) and number of levels included. 
\label{fig:CI9}
}
\end{figure}

\subsection{2D results for CI9}
For the stoichiometrical isomers C$_7$H$_{10}$O$_2$, data set CI9, we have also investigated the 2D-CQML model corresponding to Eq.~\eqref{eq:2D}.
The resulting models differ from the previous 2D-CQML models in that they unite energy approximation effects and basis sets
into a single dimension (PM7, B3LYP/6-31g(D), G4MP2).
Furthermore, and in analogy to the original $\Delta$-ML model~\cite{DeltaPaper2015,outsmartQC2018,rupp2015nmr,raghu2015excitation,tristan2017NCI,kennedy2000}, 
all small changes in geometry due to use of different level of theory, are also being accounted for through the ML model. 
As such, only PM7-quality input geometries are required for the 2D-CQML models discussed in this section.
Resulting learning curves are shown in Fig.~\ref{fig:CI9} for two different representations, 
the Coulomb matrix~\cite{RuppPRL2012,AssessmentMLJCTC2013} and SLATM~\cite{amons2017}, 
as well as for two different number of levels ($L = 2$ and $L = 3$).

Again, when compared to conventional QML, we note systematic and improved (through lower off-sets) learning as the number of different levels increases from two to three. 
The relative performance for Coulomb matrix and SLATM meets the expected trend~\cite{FCHL}, SLATM systematically leading to a substantially lower off-set.
These results suggest a certain independence of the CQML methodology from other salient features of QML models, such as training set 
selection~\cite{Nick2016GA,amons2017} 
or choice of representation~\cite{baml,FCHL}.
In this case, the best 2D-CQML SLATM based model reaches chemical accuracy with respect to G4MP2 based on a training
set consisting of $\sim$1'000, 2'000, and 4'000 at G4MP2, B3LYP/6-31g(D), and PM7 level reference results, respectively. \\
%{\color{red} PETER, IS THIS CORRECT THAT YOU USED s=1 FOR THESE RESULTS??} \textcolor{green}{Yes.}

%\caption{\label{fig:QM7_3d_s2}As in the two-space case, increasing the level scaling parameter to $s=2$ gives higher training sample reduction ratios.}
\section{Conclusions}
\label{sec:Conclusions}
%summary
%insight gained
We have extended the ideas manifested in Pople-diagrams within the systematic framework of the multi-level sparse grid combination technique and machine learning.
A generalized CQML model has been presented, and we have demonstrated its performance for various 2D variants and for one 3D application using 
atomization energies of organic molecules as property of interest.
Using learning curves to compare models, we have found for all cases investigated that the addition of levels and spaces enables
a systematic and substantial reduction in necessary training data at the highest level of theory.
As such, we have shown how to construct QML models for which an expensive training molecule can be replaced by multiple cheaper training molecules. 
Due to the unfavourable polynomial scaling and large prefactors of the more expensive quantum approximations, 
such trade-offs can deliver significantly more accurate QML models at constant training data compute budget. 
In conclusion, our numerical findings support the idea that there is an additional ``knob'' one can use to improve QML models: 
In addition to improved representations~\cite{baml,FCHL} or training set selection~\cite{Nick2016GA,amons2017}
one can also exploit the intrinsic correlations among the various hierarchies which exist among different levels of approximations.

%outlook %Multi-fidelity?
For future work, we will consider the inclusion of more intermediate levels, e.g.~the various rungs on Jacob's ladder,
or MP4, CCSD, CCSDT(Q), etc., or continuous changes in basis set size through plane-waves.
Other dimensions, such as relativistic effects, spin-orbit coupling, or nuclear quantum effects can be envisioned. 
While we have focussed on atomization energies only for this study, we will consider CQML models of other quantum properties within subsequent studies.
Technical settings can also be investigated, e.g.~the relative amount of training data obtained at different levels (currently set globally through parameter $s$), 
could still be adapted in a locally optimal manner.
Finally, we plan to include this implementation in qmlcode~\cite{qmlcode}.

\begin{appendix}
\section{Derivation of the combination technique for quantum machine learning} \label{appendix}

In applied mathematics, the sparse grid combination technique is a means to approximate, e.g., high-dimensional functions. Lets assume that such a function $f$ is in some (function) space $V := V^{(1)} \otimes V^{(2)} \otimes \cdots \otimes V^{(d)}$. That is, it is in the tensor product of $d$ spaces. Then, we introduce for each of the
{$L_m$-dimensional}
%(infinite-dimensional) 
function spaces $V^{(m)}$ a series of
subspaces of lower dimension
%(finite-dimensional) subspaces
\bea
V^{(m)}_{0} \subset V^{(m)}_{1} \subset V^{(m)}_{j} \subset \ldots \subset {V_{L_m}^{(m)}}
\eea
(indicated by the lower index).
Classic (full tensor-product) approximation would now approximate this function $f$ on a level $j$ in the space $V_{j} :=  V_j^{(1)} \otimes V_j^{(2)} \otimes \cdots \otimes V_j^{(d)}$. However, this leads to the so-called \textit{curse of dimensionality}, i.e.~the exponential growth in computational work with growing dimension $d$.

In many cases, the sparse grid combination technique allows to approximate $f$ in a much cheaper way. This is done by recursively introducing the sparse approximation space $\hat{V}_j$ with
\begin{equation}\label{eq:CTrecursive}\hat{V}_j^{(d)} := \sum_{k=0}^j \left(V_{j-k}^{(d)} - V_{j-1-k}^{(d)}\right)\otimes  \hat{V}_k^{(d-1)}\,,\end{equation}
where $\hat{V}_k^{(d-1)}$ is the sparse approximation space 
\bea 
\hat{V}_j^{(d-1)} := \sum_{k=0}^j \left(V_{j-k}^{(d-1)} - V_{j-1-k}^{(d-1)}\right)\otimes  \hat{V}_k^{(d-2)}\,.
\eea 
That is, it is recursively built from the first $d-1$ spaces in the same way. 

In this work, we transfer this approach to the field of quantum machine learning. To this end, we provide a derivation for the combination technique for quantum machine learning in two and three dimensions / spaces. Let us first briefly introduce a general machine learning model for a given subspace $(\ell_C,\ell_B,\ell_M)$. Note that we assume here that $\ell_C,\ell_B,\ell_M\in \{0,\ldots, L\}$. The general ML model for a given subspace reads as
\bea \label{eq:trans}
E_{(\ell_C,\ell_B,\ell_M)}(\vec{R}_q):=\sum_{i=1}^{N_{\ell_M}} \alpha_i^{(\ell_C,\ell_B,\ell_M)} \kernel(\vec{R}_q,\vec{R}_i)\,.
\eea 
We identify this model with some subspace $V_{\ell_C}^{(1)}\otimes V_{\ell_B}^{(2)}\otimes V_{\ell_M}^{(3)}$. 
Following equation \eqref{eq:CTrecursive}, the two-dimensional combination technique for QML on level $j_2$ for the spaces of theory and training set size and a fixed basis set level $\ell_B$ can be introduced as
\bea
\begin{split}
E_{(j_2,\ell_B,j_2)}(\vec{R}_q):= &\sum_{k_2=0}^{j_2} E_{(j_2-k_2,\ell_B,k_2)}(\vec{R}_q) \\
&- E_{(j_2-1-k_2,\ell_B,k_2)}(\vec{R}_q)
\end{split}
\eea 
Note that, whenever a level index becomes negative, we assume the machine learning model to be exactly zero, i.e.
\bea 
E_{(-1,\cdot,\cdot)}\equiv E_{(\cdot,-1,\cdot)} \equiv E_{(\cdot,\cdot,-1)}\equiv 0\,.
\eea 
For the choice of $j_2=2$ and $\ell_B=2$, we can explicitly derive
\bea
E_{(2,2,2)}(\vec{R}_q)  
  =&\left(E_{(2-0,2,0)}(\vec{R}_q) - E_{(2-1-0,2,0)}(\vec{R}_q)\right)\nonumber\\
& +\left(E_{(2-1,2,1)}(\vec{R}_q) - E_{(2-1-1,2,1)}(\vec{R}_q)\right)\nonumber\\
& +\left(E_{(2-2,2,2)}(\vec{R}_q) - E_{(2-1-2,2,2)}(\vec{R}_q)\right)\nonumber\\
= & \left(E_{(2,2,0)}(\vec{R}_q) - E_{(1,2,0)}(\vec{R}_q)\right)\nonumber\\
& +\left(E_{(1,2,1)}(\vec{R}_q) - E_{(0,2,1)}(\vec{R}_q)\right)\nonumber\\
& +\left(E_{(0,2,2)}(\vec{R}_q) - E_{(-1,2,2)}(\vec{R}_q)\right)\nonumber \\ 
= & E_{(2,2,0)}(\vec{R}_q) - E_{(1,2,0)}(\vec{R}_q)\nonumber\\
& +E_{(1,2,1)}(\vec{R}_q) - E_{(0,2,1)}(\vec{R}_q)\nonumber\\
& +E_{(0,2,2)}(\vec{R}_q)
\eea 
Note that we have the equalities 
%\textcolor{red}
{\bea %\begin{align*}
E_{(2,2,0)}(\vec{R}_q) - E_{(1,2,0)}(\vec{R}_q) &= \sum_i^{N_2}\alpha_i^{(1,2)}\kernel(\vec{R}_q,\vec{R}_i)\,,\nonumber\\
E_{(1,2,1)}(\vec{R}_q) - E_{(0,2,1)}(\vec{R}_q) &= \sum_i^{N_1}\alpha_i^{(0,1)}\kernel(\vec{R}_q,\vec{R}_i)\,,\nonumber\\
E_{(0,2,2)}(\vec{R}_q) &= \sum_i^{N_0}\alpha_i^{(0)}\kernel(\vec{R}_q,\vec{R}_i)\,,\nonumber
\eea} %\end{align*}
with the notation from Section~\ref{sec:mlqml}. 
%\textcolor{red}{WARNING: Here, we still have the issue that $N_\ell$ was switched between the 2D and the 3-nD CQML.}
That is, model $E_{(2,2,2)}$, as derived here, is exactly the model discussed in Section~\ref{sec:mlqml}.

Based on the two-dimensional combination technique model, we can now recursively build a three-dimensional combination technique further integrating the space of basis set size and with the global three-dimensional level $j_3$ as follows
\bea 
\begin{split}
E_{(j_3,j_3,j_3)}(\vec{R}_q) := &\sum_{k_3=0}^{j_3} E_{(k_3,j_3-k_3,k_3)}(\vec{R}_q) \\
&- E_{(k_3,j_3-1-k_3,k_3)}(\vec{R}_q)\,.
\end{split}
\eea 
This construction uses the definition of the two-dimensional combination technique in a recursive fashion.

We finally exemplify the tree-dimensional combination technique for $j_3=2$. That is, we first expand the recursive model for the three-dimensional combination technique by
\begin{align*}
E_{(2,2,2)}(\vec{R}_q) = &\left(E_{(0,2-0,0)}(\vec{R}_q) - E_{(0,2-1-0,0)}(\vec{R}_q)\right) \nonumber \\
& + \left(E_{(1,2-1,1)}(\vec{R}_q) - E_{(1,2-1-1,1)}(\vec{R}_q)\right) \nonumber \\
& + \left(E_{(2,2-2,2)}(\vec{R}_q) - E_{(2,2-1-2,2)}(\vec{R}_q)\right) \nonumber \\
= &\left(E_{(0,2,0)}(\vec{R}_q) - E_{(0,1,0)}(\vec{R}_q)\right) \nonumber \\
& + \left(E_{(1,1,1)}(\vec{R}_q) - E_{(1,0,1)}(\vec{R}_q)\right) \nonumber \\
& + \left(E_{(2,0,2)}(\vec{R}_q) - E_{(2,-1,2)}(\vec{R}_q)\right) \nonumber \\
= &\left(E_{(0,2,0)}(\vec{R}_q) - E_{(0,1,0)}(\vec{R}_q)\right) \nonumber \\
& + \left(E_{(1,1,1)}(\vec{R}_q) - E_{(1,0,1)}(\vec{R}_q)\right) \nonumber \\
& + E_{(2,0,2)}(\vec{R}_q)\,.
\end{align*}
Then, we expand each of the term by means of the two-dimensional combination technique. Thus we compute
\begin{align*}
E_{(0,2,0)}(\vec{R}_q) =& E_{(0-0,2,0)}(\vec{R}_q) - E_{(0-1-0,2,0)}(\vec{R}_q) \nonumber \\
=& E_{(0,2,0)}(\vec{R}_q)\,, \nonumber
\end{align*}
\begin{align*}
E_{(0,1,0)}(\vec{R}_q) =& E_{(0-0,1,0)}(\vec{R}_q) - E_{(0-1-0,1,0)}(\vec{R}_q) \nonumber \\
=& E_{(0,1,0)}(\vec{R}_q)\,,
\end{align*}
\begin{align*}
E_{(1,1,1)}(\vec{R}_q) =& E_{(1-0,1,0)}(\vec{R}_q) - E_{(1-1-0,1,0)}(\vec{R}_q) \nonumber \\
&+ E_{(1-1,1,1)}(\vec{R}_q) - E_{(1-1-1,1,1)}(\vec{R}_q) \nonumber \\
=& E_{(1,1,0)}(\vec{R}_q) - E_{(0,1,0)}(\vec{R}_q) \nonumber \\
&+ E_{(0,1,1)}(\vec{R}_q)\,,
\end{align*}
\begin{align*}
E_{(1,0,1)}(\vec{R}_q) =& E_{(1-0,0,0)}(\vec{R}_q) - E_{(1-1-0,0,0)}(\vec{R}_q) \nonumber \\
&+ E_{(1-1,0,1)}(\vec{R}_q) - E_{(1-1-1,0,1)}(\vec{R}_q) \nonumber \\
=& E_{(1,0,0)}(\vec{R}_q) - E_{(0,0,0)}(\vec{R}_q) \nonumber \\
&+ E_{(0,0,1)}(\vec{R}_q)\,,
\end{align*}
\begin{align*}
E_{(2,0,2)}(\vec{R}_q)  
  =&\left(E_{(2-0,0,0)}(\vec{R}_q) - E_{(2-1-0,0,0)}(\vec{R}_q)\right) \nonumber \\
& +\left(E_{(2-1,0,1)}(\vec{R}_q) - E_{(2-1-1,0,1)}(\vec{R}_q)\right) \nonumber \\
& +\left(E_{(2-2,0,2)}(\vec{R}_q) - E_{(2-1-2,0,2)}(\vec{R}_q)\right) \nonumber \\
= & \left(E_{(2,0,0)}(\vec{R}_q) - E_{(1,0,0)}(\vec{R}_q)\right) \nonumber \\
& +\left(E_{(1,0,1)}(\vec{R}_q) - E_{(0,0,1)}(\vec{R}_q)\right) \nonumber \\
& +\left(E_{(0,0,2)}(\vec{R}_q) - E_{(-1,0,2)}(\vec{R}_q)\right) \nonumber \\
= & \left(E_{(2,0,0)}(\vec{R}_q) - E_{(1,0,0)}(\vec{R}_q)\right) \nonumber \\
& +\left(E_{(1,0,1)}(\vec{R}_q) - E_{(0,0,1)}(\vec{R}_q)\right) \nonumber \\
& +E_{(0,0,2)}(\vec{R}_q)\,.
\end{align*}
Finally, we combine these results with the previous calculations for $E_{(2,2,2)}$ and obtain
\begin{align*}
E_{(2,2,2)}(\vec{R}_q) = & E_{(0,2,0)}(\vec{R}_q) - E_{(0,1,0)}(\vec{R}_q) \nonumber \\
& + E_{(1,1,1)}(\vec{R}_q) \nonumber - E_{(1,0,1)}(\vec{R}_q) \nonumber \\
& + E_{(2,0,2)}(\vec{R}_q) \nonumber \\
= & E_{(0,2,0)}(\vec{R}_q) - E_{(0,1,0)}(\vec{R}_q) \nonumber \\
& + \left[ E_{(1,1,0)}(\vec{R}_q) - E_{(0,1,0)}(\vec{R}_q)\right. \nonumber \\
&\quad\left. + E_{(0,1,1)}(\vec{R}_q)\right] \nonumber \\
& - \left[ E_{(1,0,0)}(\vec{R}_q) - E_{(0,0,0)}(\vec{R}_q)\right. \nonumber \\
&\quad\left. + E_{(0,0,1)}(\vec{R}_q)\right] \nonumber \\
& + \left[\left(E_{(2,0,0)}(\vec{R}_q) - E_{(1,0,0)}(\vec{R}_q)\right)\right. \nonumber \\
&\quad +\left(E_{(1,0,1)}(\vec{R}_q) - E_{(0,0,1)}(\vec{R}_q)\right) \nonumber \\
&\quad\left. + E_{(0,0,2)}(\vec{R}_q)\right] \nonumber \\
= & E_{(0,2,0)}(\vec{R}_q) - 2 E_{(0,1,0)}(\vec{R}_q) \nonumber \\
& + E_{(1,1,0)}(\vec{R}_q) + E_{(0,1,1)}(\vec{R}_q) \nonumber \\
& - 2 E_{(1,0,0)}(\vec{R}_q) + E_{(0,0,0)}(\vec{R}_q) \nonumber \\
& - 2 E_{(0,0,1)}(\vec{R}_q) + E_{(2,0,0)}(\vec{R}_q)  \nonumber \\
& + E_{(1,0,1)}(\vec{R}_q)  + E_{(0,0,2)}(\vec{R}_q)
\end{align*}
This is exactly the spelled out version of equation \eqref{eq:Example3D} for $E_{222}$.
\end{appendix}

\section*{supplementary material}
Geometries are provided as xyz files. Two types of energy data are available for each of the three basis sets (sto-3g, 6-31g and cc-pvdz), i.e., the total energy ($E$) and effective averaged atomization energies ($E^*$). The latter is defined as $E - \sum_I n_I*e_I$, where $n_I$ is the number of atom $I$ in the molecule and $e_I$ is the effective atomic energy of atom $I$ obtained through a linear least square fit of $E = \sum_I n_I*e_I$ for all molecules in the dataset.
Free atom energies for all basis sets and electron methods are also included.
%One great benefit of this definition is the circumvention of the calculation of free atoms, for which the computational accuracy is not at the same level as for molecules with saturated valence.
Every type of energy data for any basis set used is given as a text file, consisting of three columns representing HF, MP2 and CCSD(T) energies, respectively.

\begin{acknowledgements}
We are grateful for discussions with P. D. Mezei and M. Schwilk. 
This collaboration is mainly being funded by the Swiss National Science foundation through 407540\_167186 NFP 75 Big Data.
OAvL also acknowledges additional support by the Swiss National Science foundation (No.~PP00P2\_138932, 200021\_175747, NCCR MARVEL).
Some calculations were performed at sciCORE (http://scicore.unibas.ch/) scientific computing core facility at University of Basel.
\end{acknowledgements}

\bibliography{literatur,math}

\end{document}